\documentclass[fleqn,10pt]{wlscirep}
\usepackage[utf8]{inputenc}
\usepackage[T1]{fontenc}
\usepackage{lineno}
\usepackage{soul}
\usepackage{longtable}
\usepackage{multirow}
\usepackage{siunitx}
\usepackage{verbatim}
\usepackage{arydshln}
\usepackage{caption}
\usepackage{placeins}
\usepackage{makecell}
\usepackage{lscape}
\usepackage{multicol}

\newcommand{\etal}{\textit{et al.}}

\title{MedalCare-XL: 16,900 healthy and pathological 12~lead ECGs obtained through electrophysiological simulations}

\author[1,$\dag$]{Karli Gillette}
\author[1,$\dag$]{Matthias A.F. Gsell}
\author[2,$\dag$]{Claudia Nagel}
\author[2]{Jule Bender} 
\author[3]{Bejamin Winkler}
\author[4,5]{Steven E. Williams}
\author[3]{Markus B{\"a}r}
\author[3,4,6]{Tobias Sch\"{a}ffter}
\author[2,$\dag \dag$]{Olaf D{\"o}ssel}
\author[1,$\dag \dag$,*]{Gernot Plank}
\author[2,$\dag \dag$,*]{Axel Loewe}
\affil[1]{Gottfried Schatz Research Center: Division of Medical Physics and Biophysics, Medical University of Graz, Graz, Austria}
\affil[2]{Institute of Biomedical Engineering, Karlsruhe Institute of Technology (KIT), Karlsruhe, Germany}
\affil[3]{Physikalisch-Technische Bundesanstalt, National Metrology Institute, Berlin, Germany}
\affil[4]{King's College London, London, United Kingdom}
\affil[5]{University of Edinburgh, Edinburgh, United Kingdom}
\affil[6]{Biomedical Engineering, Technische Universit{\"a}t Berlin, Einstein Centre Digital Future}
\affil[*]{corresponding authors: Axel Loewe (publications@ibt.kit.edu), Gernot Plank (gernot.plank@medunigraz.at)}

\affil[$\dag$, $\dag \dag$]{these authors contributed equally to this work}

\begin{abstract}

    Mechanistic cardiac electrophysiology models allow for personalized simulations of the electrical activity in the heart and the ensuing electrocardiogram (ECG) on the body surface. As such, synthetic signals possess known ground truth labels of the underlying disease and can be employed for validation of machine learning ECG analysis tools in addition to clinical signals. Recently, synthetic ECGs were used to enrich sparse clinical data or even replace them completely during training leading to improved performance on real-world clinical test data.
    
    We thus generated a novel synthetic database comprising a total of 16,900 12~lead ECGs based on electrophysiological simulations equally distributed into healthy control and 7 pathology classes. The pathological case of myocardial infraction had 6 sub-classes. A comparison of extracted features between the virtual cohort and a publicly available clinical ECG database demonstrated that the synthetic signals represent clinical ECGs for healthy and pathological subpopulations with high fidelity. The ECG database is split into training, validation, and test folds for development and objective assessment of novel machine learning algorithms. 

\end{abstract}
\begin{document}

\flushbottom
\maketitle

\thispagestyle{empty}


\section*{Background \& Summary}

    The 12~lead ECG is a standard non-invasive clinical tool for the diagnosis and long-term monitoring of cardiovascular disease. To support cardiac disease classification and interpretation of 12~lead ECGs in clinical practice, algorithms based on machine learning are increasingly utilized. Training of these algorithms requires large databases of 12~lead ECGs that have been labeled according to desired disease classifications with high accuracy and represent the target population. The most extensive publicly available database for such purpose to date is PTB-XL~\cite{wagner2020ptb}. 
    
    Clinical 12~lead ECG databases like PTB-XL, however, have several limitations reducing efficacy of machine learning algorithms~\cite{Roberts-2021-ID15918}. As the databases are typically attained from multiple medical centers, different filtering levels may be applied to reduce noise. Labeling uncertainties may arise due to differences in expertise or judgment between clinicians. Patient enrollment can also lead to both gender bias~\cite{Puyol-Anton-2021-ID17011} and uneven representation of certain cardiac diseases~\cite{Pilia-2020-ID14680}. Furthermore, such databases provide limited insight into the underlying mechanisms of cardiovascular disease. Databases of synthetic ECGs have the potential to either complement and enrich~\cite{Luongo-2022-ID17367,Nagel-2022-ID17346}, or in the long run to even replace~\cite{Luongo-2021-ID15954}, clinical datasets to overcome such limitations.
    
     Currently, no sizeable and open synthetic ECG databases are available due to the high computational cost and limitations in modeling complete four-chamber cardiac electrophysiology \emph{in silico} at scale. While four-chamber cohorts exist for the modeling of cardiac electrophysiology \cite{strocchi2020publicly}, such cohorts are not suited for the generation of large ECG databases due to a lack of controllable electrophysiology or limited anatomical variation. Separate models of atrial and ventricular electrophysiology that are individually more detailed and steerable can be later joined together to capture the P~wave and the QRST~complex within the 12~lead ECG, respectively, and overcome such limitations. 

    We thus aimed to assemble the first public database of labeled synthetic 12~lead ECGs by joining two independent multi-scale models of atrial and ventricular electrophysiology used to compute P~waves and QRS~complexes, respectively. This approach provides a complete chain of traceability from the anatomical and electrophysiological input parameters of the model to the final 12~lead ECGs. Common diseases were modeled mechanistically in addition to normal healthy control within the synthetic database. Within the ventricular-torso model, the pathologies of myocardial infarction (MI) and complete bundle branch block of both the left ventricle (LBBB) and the right ventricle (RBBB) were modeled. The MI class comprised 6 sub-classes pertaining to the three predominant arteries of right-anterior descending (RAD), left anterior descending (LAD), and left circumflex (LCX) \cite{american2002standardized} each with two different transmural extent. The diseases fibrotic atrial cardiomyopathy (FAM), complete interatrial conduction block (IAB) and left atrial enlargement (LAE) were modeled within the atria. Also, 1st degree AV block (AVB) was modeled as an atrio-ventricular (AV) conduction-based disease. In this way, the chosen pathologies cover a wide range of both atrial and ventricular diseases representing conduction disturbances as well as structural remodeling for which established modeling approaches published in previous work could be resorted to. A total of 16,900 synthetic ECGs equally distributed into the 8 groups (healthy control and 7 cardiac pathologies) were made publicly available in the MedalCare-XL database. This MedalCare-XL dataset is publicly available under the Creative Commons Attribution 4.0 International license~\cite{medalcare-xl}. Thus, we provide a large and balanced ECG dataset with precisely known ground truth labels of the underlying pathology as derived from the mechanistic multi-scale simulations. 
        
    Validation of the synthetic ECG database was performed using two approaches to analyze to what extent the synthetic ECG database could represent clinical ECG databases. First, we tested the MedalCare-XL data set of simulated ECGs by comparing the statistical distribution of crucial ECG features extracted from MedalCare-XL with the same features taken from the clinical PTB-XL data base for normal healthy ECGs and for different pathology classes. The comparison showed excellent qualitative agreement, while still exhibiting quantitative differences that provide a starting point for future improvement of the underlying models as well as of the quality of future simulation data bases. Second, two clinical Turing tests were also conducted to evaluate the ability of the generated synthetic ECG signals to represent clinical signals undergoing ECG diagnostics by cardiologists. The first test required trained cardiologists to determine the origin of both measured and simulated 12~lead ECGs under normal healthy control. The second test additionally involved pathology classification. Both tests were performed on a subset of 50 synthetics ECG signals extracted from the database and mixed with 50 clinical signals taken from PTB-XL~\cite{wagner2020ptb}.  Altogether, the MedalCare-XL data base provides the first example for a large-scale data set of physiologically-realsitic simulated ECGs.
    
    

\section*{Methods}

\begin{figure}[h]
\includegraphics[width=\linewidth]{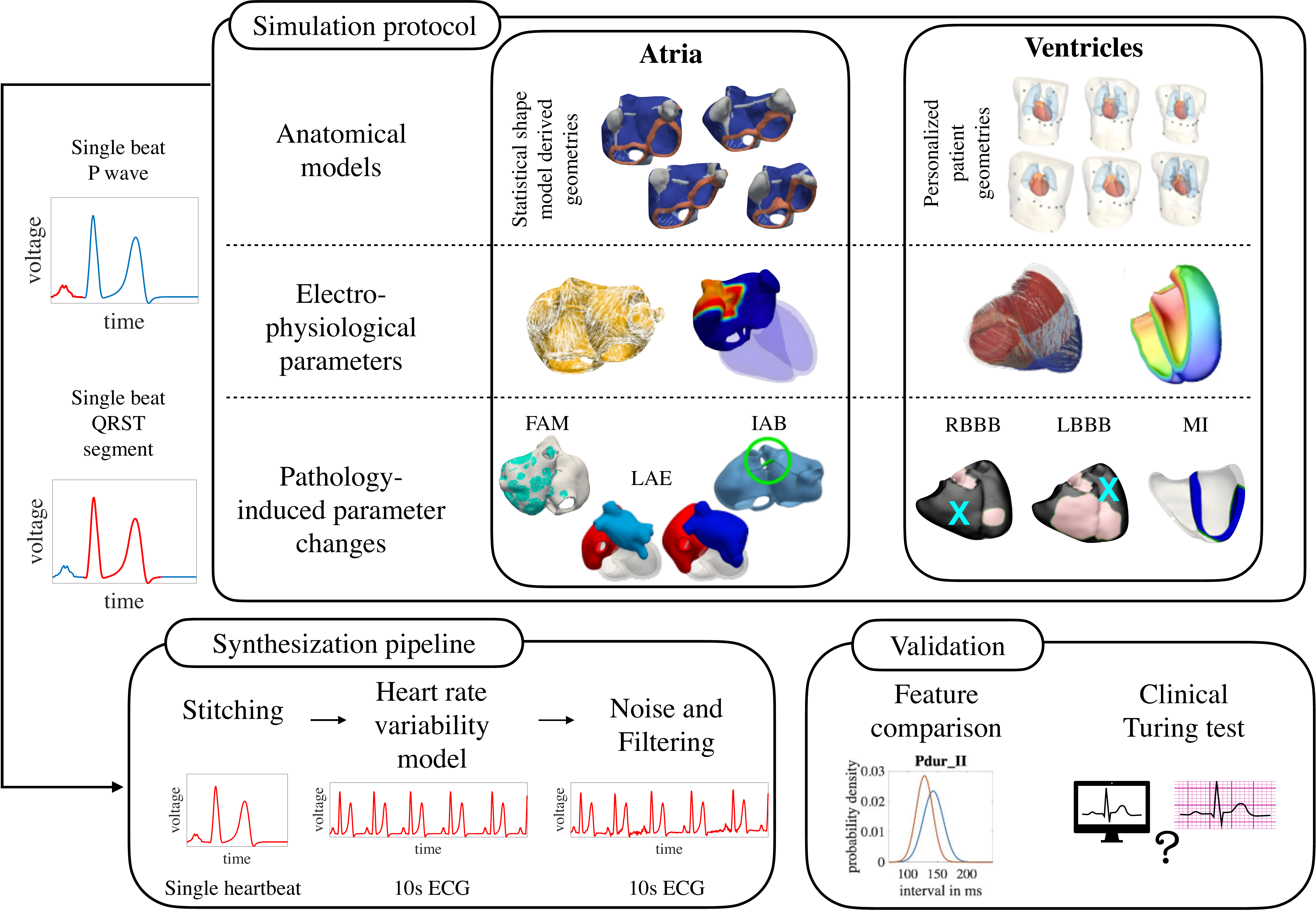}
\caption{Pipeline for the generation and validation of the synthetic 12~lead ECG database using individual multi-scale models of the atria and the ventricles.}
\label{fig:pipeline}
\end{figure}

    We separate the genesis of the 12~lead ECG into P~waves and the QRST~complex, modeled by two separate atrial and ventricle-torso models. Generation of the anatomical model cohorts and the simulation of electrophysiology to mimic a large patient population is described for both the atrial and ventricular models. Having run single beat simulations for P~waves and QRST~complexes separately in the two independent models, both signal parts had to be merged in a post-processing step to obtain an ECG of a full heart cycle comprising one P~wave, one QRS~complex and one T~wave. Subsequently, the single heartbeat was repeated with varying RR intervals to account for heart rate variability (HRV) to obtain a time series signal of 10\,$\si{\second}$ length. 
    A visual overview of the pipeline for generating the synthetic 12~lead ECG database is visualized in Figure~\ref{fig:pipeline}. 
    
    \subsection*{Anatomical Model Populations}
    
        \subsubsection*{Ventricles}
        
\begin{figure}[h]
\includegraphics[width=\linewidth]{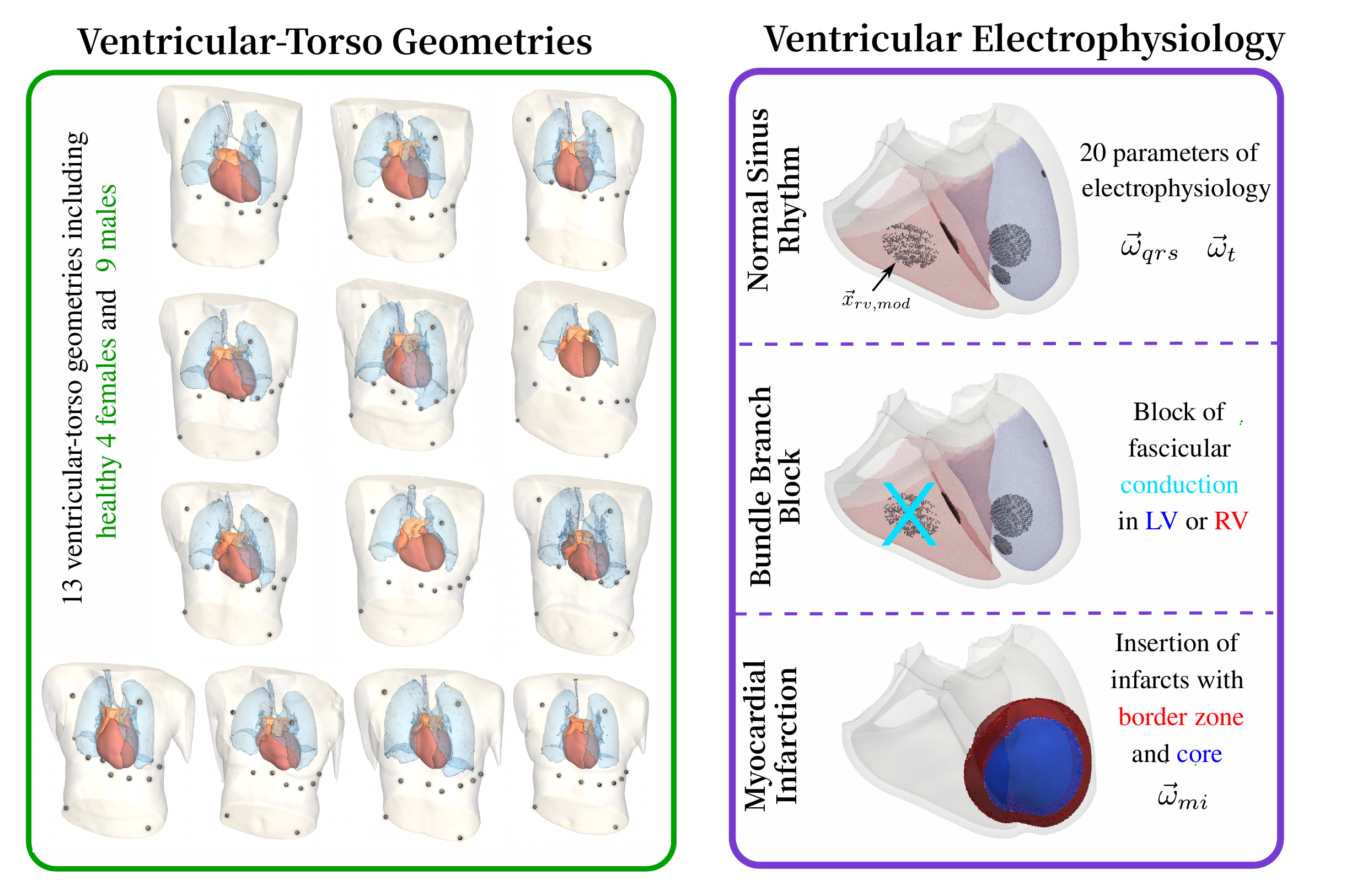}
\caption{Cohort of ventricular-torso models derived from clinical MRIs. Tissues include lungs, blood pools, atrial tissue, ventricles, and general torso. Parameters dictating ventricular electrophysiologyfor normal healthy control were varied through physiological ranges. Disease conditions of BBB and MI were then modeled by making adaptions to the model. }
\label{fig:ven_cohort}
\end{figure}

            A cohort of anatomically-specific ventricular-torso models was generated for 13 healthy subjects (8~M, 5~F) ranging from 30 to 65 years of age. All subjects were part of a clinical study approved by ethical review board at the Medical University of Graz (EKNr: 24–126 ex 11/12). Written and informed consent for each subject was attained at the time of the study. Two separate MRI scans of the full torso and whole heart were sequentially acquired using standardized protocols at 3T (Magnetom Skyra, Siemens Healthcare, Erlangen, Germany). The torso MRI (1.3 x 1.3 x 3.0 $\si{\milli \meter \cubed}$) was acquired in four overlapping stacks using a non-ECG gated 3D T1-weighted gradient-echo sequence. The whole heart MRI (0.7 x 0.7 x 0.7 $\si{\milli \meter \cubed}$) was acquired using an ECG-gated, fat-saturated, T2-prepared, isotropic 3D gradient-echo sequence. Respiratory navigators were employed to gate the MR-acquisition under free-breathing to end-expiration. MRI-compatible electrodes for recording the 12~lead ECG of each subject were left intact during image acquisition. 
            Intensity thresholding techniques implemented in \emph{Seg$3D$}~\cite{SCI:Seg3D} were used to segment each torso MRI into heart, lungs, and general torso tissue. Segmentation of the cardiac MRI was automatically performed using a two-kernel convolutional neural network. The network was tailored for MRIs from the original network implemented for computed tomography images~\cite{payer2017multi}. Segmented structures included blood pools, ventricles, and general atrial tissue. To automatically register the four-chamber heart segmentation into the torso, an iterative closest point algorithm was utilized in \emph{Seg$3D$}~\cite{SCI:Seg3D,chetverikov2002trimmed}. 
            Anatomical meshes were generated automatically from the joint segmentations using the Tarantula software meshing package \cite{prassl2009automatically}. Target resolutions within the cardiac and torso surfaces of \SI{1.2}{\milli \meter} and \SI{4.0}{\milli \meter} were prescribed, respectively. All models within the cohort were equipped with universal ventricular coordinates (UVCs) to allow for automated manipulation of all geometric-based entities~\cite{gillette2021framework, bayer2018universal}. The entire framework for the generation of the ventricular-torso model cohort is described in detail in Gillette~\etal\cite{gillette2021framework}. The ventricular-torso model cohort comprising geometries  $\Gamma_{V,i}, i \in [1,13]$ is visualized in Figure~\ref{fig:ven_cohort}.
    
        \subsubsection*{Atria}
        
\begin{figure}[ht]
\centering
\includegraphics[width=\linewidth]{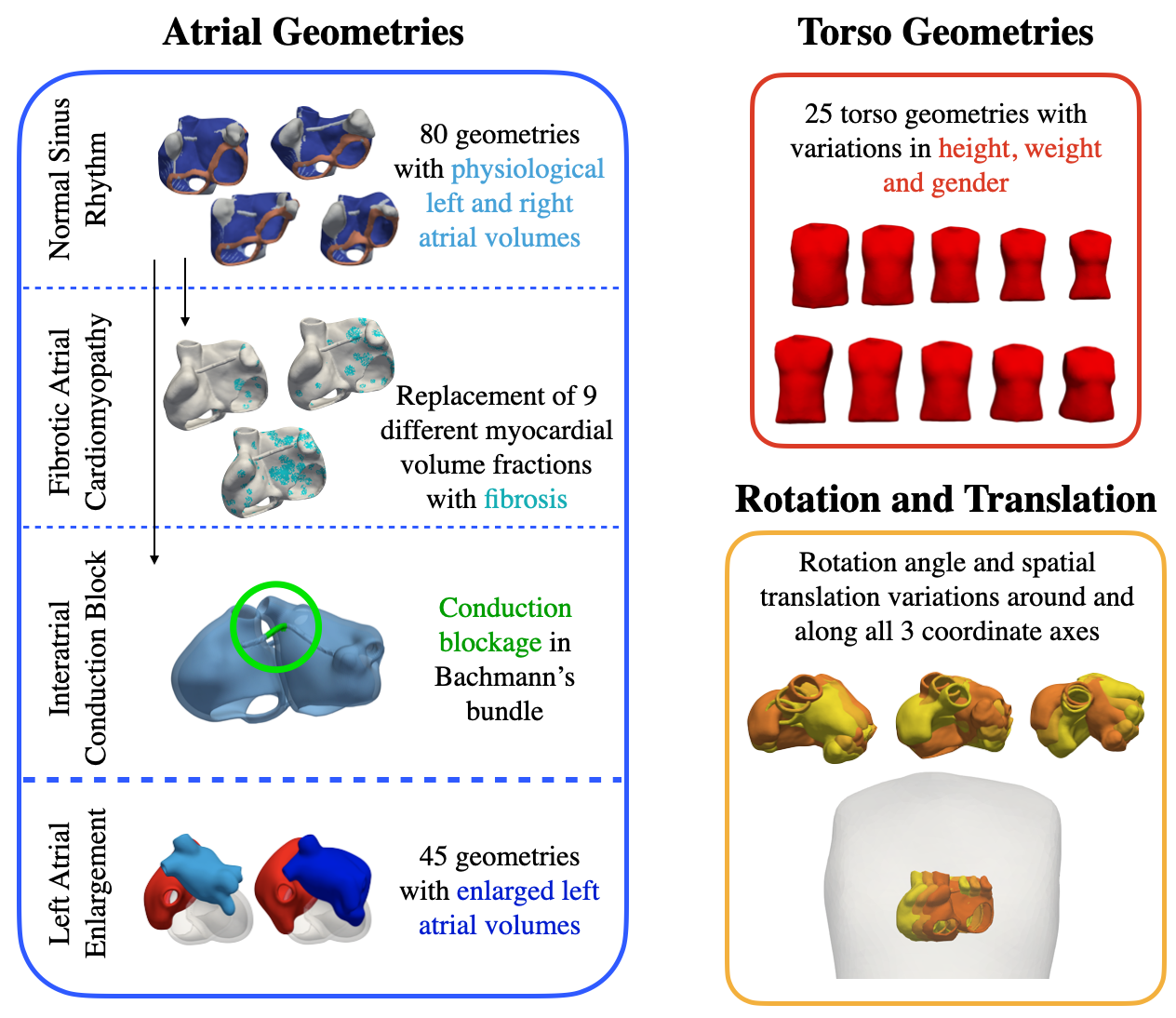}
\caption{Anatomical model cohort for atrial simulations. 80 atrial geometries with physiological left and right atrial volumes were derived from a bi-atrial statistical shape model~\cite{Nagel-2021-ID16581} and served as a basis for normal healthy control simulations. 9 different volume fractions of these models were additionally replaced by fibrosis for simulations of fibrotic atrial cardiomyopathy. Interatrial conduction block signals were generated by blocking conduction in Bachmann's Bundle in the same 80 geometries. Furthermore, 45 geometries with enlarged left atrial volumes were generated. As for the torso anatomy, 25 geometries were derived from a human body statistical shape model to account for height, weight and gender differences in the virtual patient cohort. Moreover, the rotation angle as well as the spatial position of the atria inside the torso were varied in physiological ranges.}
\label{fig:overviewAtrialAnatomicalModels}
\end{figure}

            An overview of the anatomical model cohort generated for the atrial simulations is shown in Figure~\ref{fig:overviewAtrialAnatomicalModels}. A total of 125 anatomical models $\Gamma_{A,h,i}, i \in [1,80]$ and $\Gamma_{A,LAE,i}, i \in [1,45]$ of the atrial endocardium were derived from a bi-atrial statistical shape model~\cite{Nagel-2021-ID16581, nagel_claudia_2020_4309958}. The endocardial surfaces were augmented with a homogeneous wall thickness of \SI{3}{\milli \meter}, rule-based myocardial fiber orientation, tags for anatomical structures and interatrial connections as described by Azzolin~\etal\cite{Azzolin-2022-ID17490,Zheng-2021-ID17989}. Out of these 125 geometries, 80 models exhibited left and right atrial volumes in physiological ranges reported for healthy subjects~\cite{Lang-2015-ID16479}. In these geometries, 10 different fractions from 0 to 45\,\% of the atrial myocardial tissue volume were additionally replaced by fibrotic patches as described previously~\cite{Nagel-2021-ID16114} to model atrial cardiomyopathy. The remaining 45 anatomical models were generated by constraining the coefficients of the statistical shape model such that left atrial volumes were increased to value ranges typically observed in left atrial enlargement patients~\cite{Lang-2015-ID16479}. Additionally, 25 torso geometries $\Gamma_{T,i}, i \in [1,25]$ were obtained by modifying the coefficients of the two leading eigenmodes in the human body statistical shape model constructed by Pishchulin~\etal~\cite{Pishchulin-2017-ID12226}. In this way, height, weight and gender differences were represented in the anatomical torso model cohort. By applying random rotation angles $\alpha_x, \alpha_y, \alpha_z$ and translation parameters $t_x, t_y, t_z$ in ranges summarized in Table~\ref{tab:atria_params} to the atrial geometry, heart location and orientation variability were additionally accounted for in the virtual population.

    \subsection*{Simulation Protocol and Parameters}
    
        \subsubsection*{Ventricles}
        
            Under normal healthy control, activation of the ventricles was assumed to be Durrer-based~\cite{durrer1970total}, where the His-Purkinje System was modeled assuming 5 fascicular sites of earliest breakthrough on a fast-conducting endocardium. Three fascicular sites were placed in the left ventricle (LV) on the anterior endocardium $\vec{x}_{lv,ant}$, posterior endocardium $\vec{x}_{lv,post}$, and the septum $\vec{x}_{lv,sept}$. Activation of the right ventricle (RV) was controlled using a site corresponding to the moderator band $\vec{x}_{rv,mod}$. 
            An additional site $\vec{x}_{rv,sept}$ was also placed on the right-ventricular septum. All fascicular sites were defined in UVCs. The RV moderator band was placed in the middle of the RV free wall. The transmural depth of the remaining fascicular sites was assumed to be constant at $20\,\%$ of the ventricular free wall. The fascicles were assumed to be of disc-like shape with a transmural thickness of $0.5\,\%$ of the ventricular wall, and a radius controlled through additional parameter $\vec{r}$ that related to endocardial extent. Activation was assumed to be simultaneous, apart from a prescribed delay $\vec{t}_{mod}$ in the activation of the RV moderator band site. 
        
            To modulate the fast spread of conduction on the endocardial surface of the ventricles modulated by the His-Purkinje System, a fast-conducting endocardium was also included that spanned from the middle $10\,\%$ to $90\,\%$ of the ventricular mesh along the apico-basal direction. Details of the His-Purkinje representation are available in Gillette~\etal~\cite{gillette2021framework}. An isotropic conduction velocity of $\SI{2.0}{\meter \per \second}$ was prescribed within the fast-conducting endocardium~\cite{kassebaum1966electrophysiological}.
            
            Myocardial fiber directions were applied using a rule-based method~\cite{bayer2012novel} that assumed principal fiber directions rotate radially from $60.0^\circ$ on the endocardium to the epicardium $-60.0^\circ$~\cite{streeter1969fiber}. Corresponding sheet fiber directions of $-65.0^\circ$ and $25.0^\circ$ were applied, respectively~\cite{streeter1969fiber}. Conduction velocity along the principal direction of myocardial fibers of $\SI{0.6}{\meter \per \second}$ was applied with an off-axis conduction velocity ratio of 4:2:1~\cite{taggart2000inhomogeneous}. Conductivity within the myocardium was set according to Roberts~\etal~\cite{roberts1982effect}. All remaining conductivities within the volume conductor containing lungs, blood pools, atria, and general torso tissue were set according to Keller \etal~\cite{keller2010ranking}. 
            
            Ventricular myocyte electrophysiology was modeled using the Mitchell-Schaeffer ionic model $\vec{i}_{sinus}$~\cite{mitchell2003two}. A resting membrane voltage of $\SI{-86.2}{\milli \volt}$ and a peak action potential voltage of $\SI{40}{\milli \volt}$ was assumed. Gradients in action potential duration (APD) within the myocardium, needed to establish physiological T~waves, were generated by utilizing a known relationship between the $\tau_{close}$ parameter and APDs. A linear combination of the UVCs weighted with given weights $\vec{q}_w$ was first computed at each node of the mesh. The weighted UVC gradients were mapped into a range between $APD_{min}$ and $APD_{max}$ to generate an APD map within the entirety of the ventricles.   Values for the gradients and the APD are derived from the literature~\cite{opthof2017cardiac,opthof2016dispersion,keller2011influence}. In total, variation in electrophysiology during normal healthy control was controlled through 20 variable parameters summarized in the parameter vector $\vec{\omega}_{qrs}$ for the QRS~complex:
            \begin{align}
            \vec{\omega}_{qrs} = \{ \ \vec{x}_{lv,ant}, \ \vec{x}_{lv,post}, \ \vec{x}_{lv,sept}, \ \vec{x}_{rv,mod}, \ \vec{x}_{rv,sept}, \ \vec{t}_{mod} \ \}
            \end{align}
            and $\vec{\omega}_{t}$ for the ~T~wave:
            \begin{align}
            \vec{\omega}_{t} = \{ \ \vec{i}_{sinus}, \ APD_{min}, \ APD_{max}, \ \vec{q}_w \ \}.
            \end{align}
            
            All geometric-based parameters could be mapped into the mesh using $k$D-trees implemented in \emph{meshtool}~\cite{neic2020meshtool}. Parameters relating to both the QRS~complex and T~wave under normal healthy control were varied in physiological ranges to generate variation in the QRST~complex as reported in Table~\ref{tab:qrs_params} and Table~\ref{tab:twave_params}, respectively. Sampling through the ranges for each of the parameters was done using Latin Hyper Cubes. 

            The two pathologies of BBB and MI were then modeled in the ventricles alongside normal healthy control. Pathologies of LBBB and RBBB were included in the ventricular-torso model. To cause a complete branch block, all fascicular root sites within either the LV or the RV were neglected to inhibit activation. All other relevant electrophysiology parameters were allowed to vary in the same ranges as reported for normal healthy control above. 
            
            A MI stemming from occlusion of one of the three primary arteries of RAD, LAD, and LCX was inserted into the ventricles. For each of the arteries $\nu \in \{RAD,LAD,LCX\}$, a core center $\vec{x}_{\nu,mi}$ was defined using the apico-basal and rotational UVC coordinate values that were bounded according to recommendations of affected regions on the clinical 17-segment model determined by the American Heart Association (AHA) \cite{american2002standardized}. Namely, the LAD was restricted to the anterior-anteroseptal region spanning the entire apico-basal extent. Both the RAD and LCX extended less apically, and were confined to the lateral wall and the inferior-inferioseptal regions, respectively. For each artery, the infarct was either assumed to span the entirety of the ventricular wall or  transmural extent of $30\%$ from the endocardium, giving rise to a transmural extent value $\rho_{n,mi}$ such that $n \in \{0.3,1.0\}$. The outer $5\,\%$ of the infarct area was allocated to be border zone (BZ), and the remaining area was defined as the infarct core. All scars were assumed to be left-sided, thus presenting only in LV. 
            
            From each infarct center, an Eikonal activation map was computed within the ventricular geometry assuming the same conduction velocity and off-axis ratios as assigned in the general myocardium during normal healthy control. An infarct geometry was taken by thresholding the activation map according to the computed time that generated a radius of distance $d_{co}$. The infarct core was assumed to be electrically inert, while the conduction velocity in the BZ was set to $\SI{0.15}{\meter \per \second}$ with an off-axis ratio of 1.0~\cite{mendonca2018modeling}. The conductivity within the BZ was set to the same values reported for the healthy myocardium. Parameters for the Mitchell-Schaeffer ionic model within the BZ $\vec{i}_{bz}$ were manually adjusted using bench leading to characteristic action potential changes during MI ~\cite{loewe18}. In total, the MI class comprised 6 sub-classes. The parameters varied to induce various degrees and positions of MI $\vec{\omega}_{\nu,mi}$ included: 
                \begin{align}
                \vec{\omega}_{mi} = \{ \ \vec{x}_{\nu,mi}, \rho_{n,mi} \ d_{co} \ \} : \ \nu \in \{RAD,LAD,LCX\}, \ n \in \{0.3,1.0\} 
                \end{align}
            Parameters were similarly varied using Latin Hyper Cubes through ranges based on clinical observation for characteristic occlusion sites and action potential changes 
            (Table~\ref{tab:mi_params}). 
        
            Transmembrane voltages were simulated using the efficient reaction-Eikonal method in the monodomain formulation without diffusion~\cite{neic2017reaction_eikonal}. Electrical potentials of each electrode on the torso surface were recovered from transmembrane voltages using lead fields precomputed once for every model~\cite{potse2018scalable}. A ventricular 12~lead ECG (QRST~complex) was generated by simulating a ventricular beat for $\SI{450}{\milli \second}$. All simulations were run using the \emph{CARPentry} cardiac solver~\cite{vigmond2008solvers} and the \emph{openCARP} simulation framework~\cite{plank2021opencarp,openCARP_sw} on a desktop machine with 24 cores, parallelized into 3 threads. 
    
        \subsubsection*{Atria}
        
            Local activation times in the atria were obtained by solving the Eikonal equation with the Fast Iterative Method~\cite{Fu-2013-ID14300} and the Fast Marching Method~\cite{Loewe-2019-ID12386}. Excitation was initiated at the sinoatrial node with an exit site located at the junction of crista terminalis and the superior vena cava. Locally heterogeneous conduction velocity $\mathrm{CV_{[Region]}}$ and anisotropy ratios $\mathrm{AR_{[Region]}}$ for $\mathrm{[Region]} \in \mathrm{\{bulk
           ~tissue, interatrial~connections, crista~terminalis, pectinate~muscles, inferior~isthmus\}}$ were modeled as summarized in Table~\ref{tab:atria_params}. The spatio-temporal distributions of transmembrane voltages $\mathrm{TMV}(t,x)$ were subsequently derived from the computed activation times by shifting pre-computed Courtemanche~\etal~action potential templates $\mathrm{TMV}(t)$ in time. Remodeling of cellular electrophysiology was applied in fibrotic regions as described below. For all simulations except for those of fibrotic atrial cardiomyopathy, the baseline parameters of the Courtemanche~\etal~model remained unchanged in all atrial regions.
            The atria were placed inside a torso geometry and were rotated ($\alpha_x, \alpha_y, \alpha_z$) and translated ($t_x, t_y, t_z$) around and along all three coordinate axes to account for additional anatomical variability in the cohort. 
            The forward problem of electrocardiography was solved with the infinite volume conductor method (for the normal healthy control cases and fibrotic atrial cardiomyopathy) or the boundary element method (for interatrial conduction block and left atrial enlargement). Single beat 12~lead ECGs of the P~wave lasting $\SI{150}-\SI{200}{\milli \second}$ were subsequently extracted at standard electrode positions. In total, variation during healthy sinus rhythm simulations was controlled through the parameters summarized in the following vector
            \begin{equation}
                \omega_{P} = \{ \vec{\mathrm{CV}}_{[Region]}, \alpha_x, \alpha_y, \alpha_z, t_x, t_y, t_z, \vec{\lambda}_{T,i}, \vec{\lambda}_{A,i},\}.
            \end{equation}
            
            For simulations of fibrotic atrial cardiomyopathy, nine different fractions from 5\,\% to 45\,\% of the healthy atrial myocardial volume were replaced by fibrotic tissue as described in detail by Nagel~\etal~\cite{Nagel-2021-ID16114} in the same 80 atrial anatomical models that were employed for the healthy control simulations. 
            In fibrotic patches, 50\,\% of the cells were modeled as passive conduction barriers by removing the affected elements from the volumetric meshes. In the remaining 50\,\% of the fibrotic cells, conduction velocity was reduced by a factor of 0.2 and 0.5 compared to the healthy baseline values in Table~\ref{tab:atria_params} in transversal and longitudinal fiber direction, respectively. In this way, anisotropy ratios were increased by a factor of 2.5, which typically facilitates functional reentry in patients with atrial fibrillation. To account for paracrine cytokine remodeling effects in fibrotic regions, maximum ionic conductances of the Courtemanche~\etal~cell model were rescaled (0.6$\times g_{Na}$, 0.5$\times g_{K1}$, 0.5$\times g_{CaL}$). 
            
            For left atrial enlargement simulations, 45 additional atrial geometries were derived from the bi-atrial statistical shape model. Constraints were applied to the coefficients of the leading eigenmodes to generate anatomical atrial models with systematically increasing left atrial volumes~\cite{Nagel-2022-ID17346}. Different rotation angle combinations and conduction velocity variations were applied for the simulations as reported in Table~\ref{tab:atria_params}.
            
            Complete interatrial conduction block was modeled by inhibiting conduction propagation through the elements in Bachmann's bundle at the junction between the left and the right atrium in the same 80 bi-atrial geometries that were used for the control simulations.
            Different combinations of rotation angles and spatial translations of the atria within the torso were applied for the ECG calculations. 

    \subsection*{Synthesization of Complete ECGs}
    
        Signal components were synthesized to a full ECG using a heart rate variability (HRV) model to obtain 10\,s recordings in accordance with the standard clinical 12~lead ECG. 
        As atrial and ventricular ECGs were carried out using different forward calculation methods, the amplitudes of P~waves and QRST~complexes needed to be scaled prior to concatenation to ensure that signal amplitudes of single waveforms are consistent within one heartbeat. Thus, maximum P~wave and R~peak amplitudes were extracted in lead II of all clinical recordings from healthy subjects in PTB-XL~\cite{wagner2020ptb} using ECGdeli~\cite{Pilia-2021-ID15608}. Based on these values, a multi-variate normal distribution was set up representing the relation between P~wave and R~peak amplitudes in clinical ECGs. 
        In this way, the simulated QRST~complex could be scaled with a factor sampled from this multi-variate probability distribution to match the amplitude of the simulated P~wave. A PQ~interval complying with the simulated P~wave duration was selected like-wise by drawing from a multi-variate normal distribution generated from clinical P~wave duration and PQ~interval values. Finally, the P~waves and the scaled QRST~complexes were concatenated using a sigmoid shaped segment of a length determined by the difference of PQ~interval and P~wave duration. When synthesizing ECG segments for the 1st degree AV block class, the PQ~interval was sampled from the range $>$200\,$\si{\milli \second}$.
        
        To account for heart rate variability in the simulated 10\,$\si{\second}$ ECGs, we refrained from simply repeating the concatenated single heart beat multiple times. Instead, the heart rate variability model developed by Kantelhardt~\etal~\cite{Kantelhardt_2003} was used to generate a series of RR~intervals for an average heart rate within physiological ranges (50-90\,bpm) determined from the QT~interval of the respective simulation run using the multi-variate normal distribution. For each heart beat holding a different RR~interval, the signal was shrunk or stretched in the [QRS$_\mathrm{off}$, T$_\mathrm{off}$] interval, again by sampling values from a multi-variate normal distribution derived from clinical QRS~duration, QT- and RR~interval values. After adding a sigmoidal shaped TP~segment to connect subsequent heart beats in the defined RR~interval, we obtained the final 10\,$\si{\second}$ 12~lead ECG. The raw ECG signal was superimposed with realistic ECG noise as reported by Petranas~\etal~\cite{Petrenas-2017-ID12969}. The amplitudes of the noise vectors were scaled based on a chosen signal to noise ratio between 15 and 20\,dB.
    

    
    
    
    

\section*{Data Records}

\begin{landscape}
\begin{figure}
    \centering
    \includegraphics[width=1.0\linewidth]{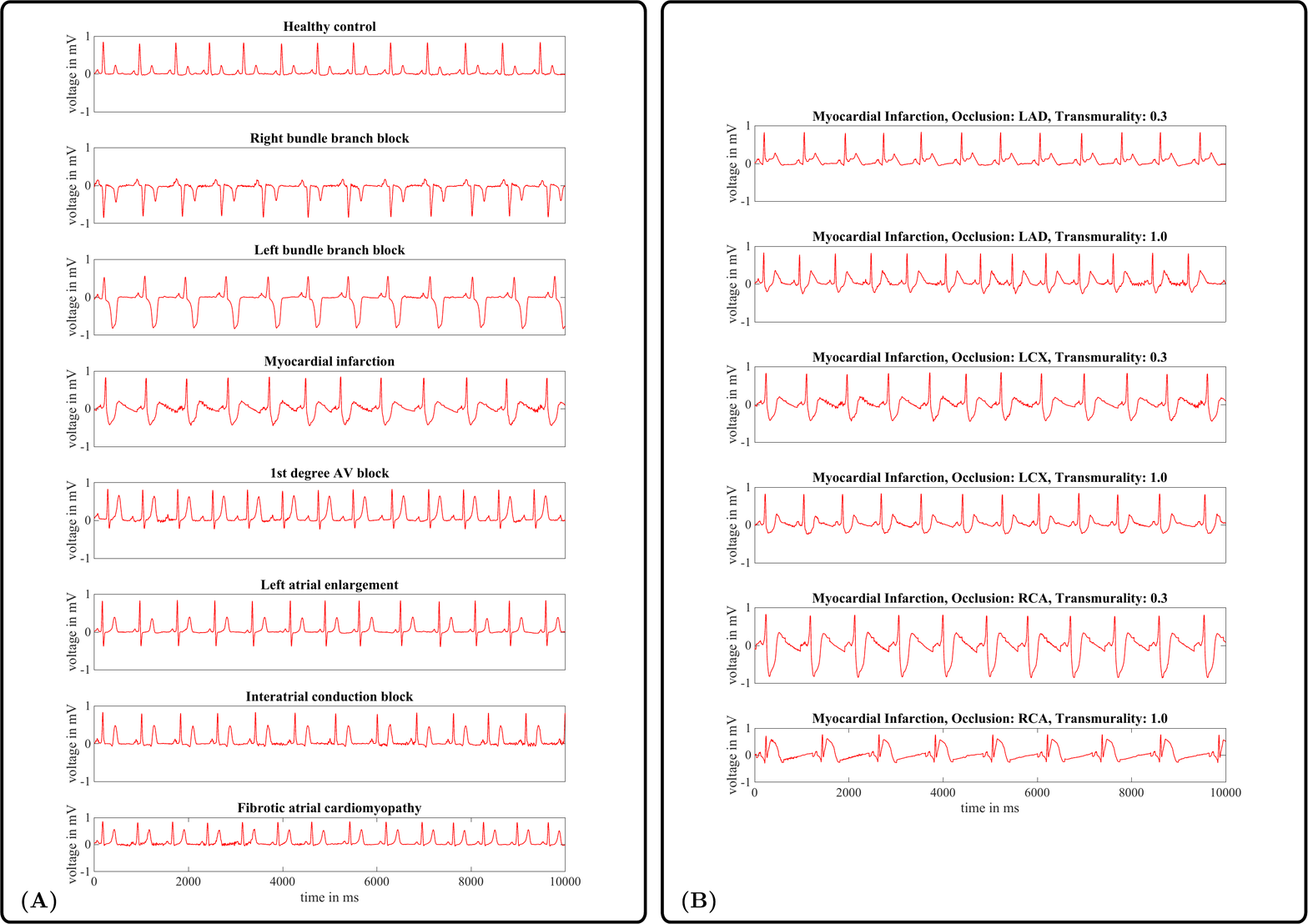}
    \caption{\textbf{(A)} Exemplary 10\,$\si{\second}$ ECGs (lead II) of each 
             pathology class and a normal healthy control in the virtual cohort.
             \textbf{(B)} Exemplary 10\,$\si{\second}$ ECGs (lead II) of each MI
             pathology class for different occlusion sites and degrees of
             transmurality.}
    \label{fig:example_ECGs}
\end{figure}
\end{landscape}

    The MedalCare-XL dataset is publicly available under the Creative Commons Attribution 4.0 International license~\cite{medalcare-xl}.
    Approximately 1,300 ECGs of 10$\si{\second}$ length for each disease class are stored in csv format. Rows 1-12 contain the 12~leads of each ECG following the order I, II, III, aVR, aVL, aVF, V1-V6. All signals are sampled at 500\,$\si{\hertz}$, amplitudes are in $\si{\milli \volt}$. Each signal is available in three different versions: 'run\_*\_raw.csv' contains the noise-free synthesized ECG, 'run\_*\_noise.csv' contains the synthesized ECG with superimposed realistic ECG noise~\cite{Petrenas-2017-ID12969}, 'run\_*\_filtered.csv' contains the bandpass filtered version (Butterworth filters of order 3, cut off frequencies of 0.5\,$\SI{}{\hertz}$ (highpass) and 150\,$\SI{}{\hertz}$ (lowpass)) of the synthesized ECGs with superimposed noise. For meaningful machine learning approaches, the signals are split in suggested subsets for training, validation and testing depending on the atrial and ventricular anatomical models the single simulation runs were based on to make sure each anatomical model is only contained in one of the subsets. Example ECGs of lead II for each disease are shown in Figure~\ref{fig:example_ECGs}~{(A)}. In Figure~\ref{fig:example_ECGs}~{(B)}, exemplary ECGs for each MI pathology class are shown corresponding to different occlusion sites and degrees of transmurality. 
    
    

\section*{Technical Validation}
We have employed two different approaches for the technical validation of the MedalCare-XL dataset of simulated, synthetic 12~lead ECGs as described in the following. For a validation of the complete dataset, the statistical distribution of ECG features extracted separately for each class (healthy control and specific pathologies) from the records in the MedalCare-XL database were compared to the distributions of the corresponding features extracted from the clinical PTB-XL that were recently summarized in the PTB-XL-Feat dataset~\cite{PTB-XL-Plus}. In addition, we performed several so-called clinical Turing tests, where the ability of expert cardiologists to distinguish the simulated ECGs from clinical ECGs was evaluated again with  representative samples from the MedalCare-XL and PTB-XL databases as described in detail below.

    \subsection*{Feature Distribution}
\begin{figure}[ht]
\centering
\includegraphics[width=0.8\linewidth]{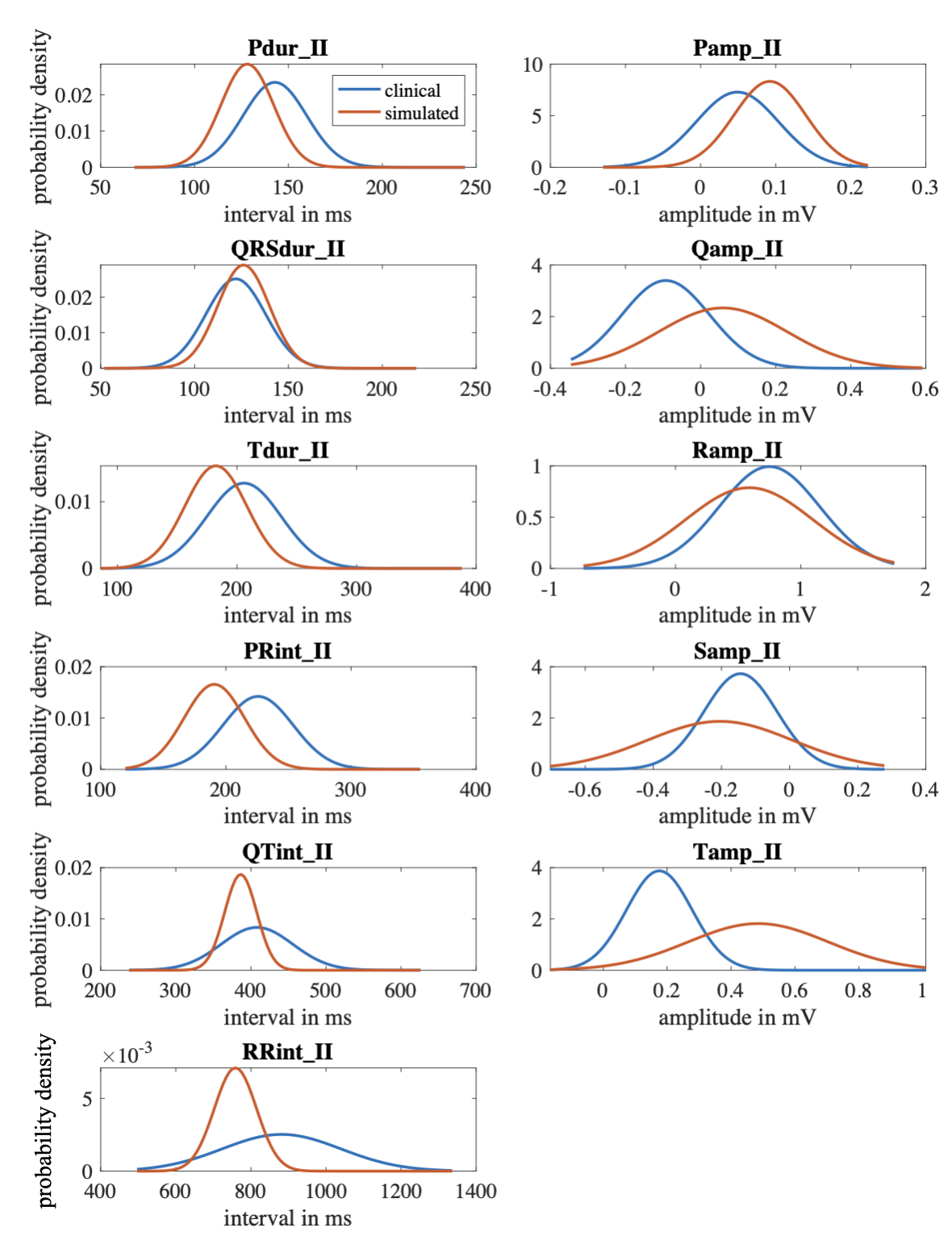}
\caption{Comparison of features in the healthy clinical and virtual cohort. Probability density functions are shown for timing features (left column, from top to bottom: P~wave duration, QRS~duration, T~wave duration, PR~interval, QTinterval, RR~interval) and amplitude features (right column, from top to bottom: P~wave amplitude, Q~/~R~/~S~peak amplitude, T~wave amplitude). Blue and red curves represent the distributions calculated based on the clinical and the simulated data, respectively. }
\label{fig:comparisonFeaturesHealthy}
\end{figure}

        To validate the simulated data against the statistical properties of clinically recorded ECGs, interval and amplitude features were extracted from the synthetic dataset and from PTB-XL using ECGdeli~\cite{Pilia-2021-ID15608} and compared to one another. Figure~\ref{fig:comparisonFeaturesHealthy} shows the probability density functions for 6 timing and 5 amplitude features extracted from lead II of all ECGs in the healthy clinical and virtual cohort. Except for the T~wave amplitudes, the feature values for the synthetic signals lie within the clinical and physiological ranges. However, the feature distributions from the healthy and the virtual data do only coincide for the QRS~duration. All other simulated timing and amplitude features only cover a subset of the clinically observed ranges. 
        
        In Figure~\ref{fig:comparisonFeaturesDiseases}, a comparison of feature distributions for healthy and pathological ECGs in the virtual cohort (top panel) and the clinical cohort (bottom panel) is visualized for timing or amplitude features that are clinically considered for a diagnosis of the respective disease. It is apparent that the change in feature values extracted from healthy and diseased ECGs is consistent between the simulated and the clinical data even though absolute feature ranges sometimes deviate. 
        
    \subsection*{Clinical Turing Tests}
    
        We aimed to ensure that the synthetic ECG signals correspond to the clinically measured 
        signals with respect to ECG features which are characteristic for healthy cases. If cardiologists are 
        not able to distinguish between measured and simulated ECG signals, this will increase confidence in 
        the \emph{in-silico} model as a surrogate for real clinical data. Therefore such a test can be considered as a clinical Turing test. For this,  cardiologists were asked to perform an online Turing test to evaluate
        and to provide feedback on both healthy and pathological ECGs. A first clinical Turing test was conducted to determine the ability of the 
        synthetic 12~lead ECGs within the database to pass as real clinical signals.
        In a second test, cardiologists were asked to determine the pathology of the signals as conducted routinely in ECG diagnostics. 
        Under all clinical Turing tests, the PTB-XL \cite{wagner2020ptb} database served as the basis for the measured signals and the simulated database described above was used for the synthetically generated
        signals.

    \subsubsection*{Development of Online Platform for Clinical Turing Test}
    
            In order to conduct clinical Turing tests,  
            an online solution provided by the \href{https://www.know-center.at}{Know-Center\footnote{https://www.know-center.at}}, 
            a research center for data science and artificial intelligence located in Graz, was used. The Know-Center extended its 
            \href{https://ecgviewer.timefuse.io/public/login/turing}
            {TimeFuse\footnote{https://ecgviewer.timefuse.io/public/login/turing}}
            online signal data platform to include a survey feature and a plotter
            to visualize 12~lead ECG signals. The ECG plotter was designed specifically to present 12~lead ECGs in a typical visualization as seen by cardiologists in the clinic on chart paper. Namely, horizontal lines on the pink background correspond to \SI{0.4}{ \second} and vertical lines correspond to \SI{0.1}{\milli \volt}. The platform was also designed for hosting of multiple clinical Turing tests. Clinical Turing tests of either healthy signals or pathological signals could then be organized and conducted separately. 
            
    \subsubsection*{Conducting Tests}

             In a first iteration, Turing tests were performed with normal healthy control ECGs to better understand the ability of signals to pass as clinical signals under normal healthy. For this
            purpose, five groups with 20 signals each were created, resulting in a total 
            of 100 signals. For the measured ECGs, 50 signals were randomly selected from
            a subset of the PTB-XL database, which contained only signals annotated as 
            100\% healthy. For the generated ECGs, 50 signals under healthy sinus rhythm 
            were randomly taken from the synthetic database described above. After pre-processing and filtering the 100 signals, the five groups were uploaded to the online platform 
            and assigned to the survey participants. Within the test, expert cardiologists were required to evaluate whether each ECG test case from the total 100 was measured or generated. Clinicians were also allowed to refrain from answering, but a lack of a statement was taken as a false classification. All clinicians were also asked to provide reasoning behind the classification. A total 
           of 6 clinicians performed the test.

           A similar test was also performed with pathological conditions to demonstrate that the synthetic ECGs of the various modeled pathological cases would be classified by expert clinicans at the same accuracy as real clinical signals and could not be distinguished from clinically
           measured ECGs taken from the PTB-XL database. The cases included myocardial infraction (MI), left bundle branch block (LBBB), 
           right bundle branch block (RBBB), first degree AV block (1AVB), and left atrial overload/enlargement (LAO/LAE). Conditions of fibrotic atrial cardiomyopathy (FAM) and complete interatrial conduction block (IAB) were neglected as such diseases were not present within PTB-XL. Examples of the disease are provided in Figures~\ref{fig:example_ECGs}~{(A)} and \ref{fig:example_ECGs}~{(B)}.

           Similar to the healthy Turing test, 50 generated ECG signals were taken from the synthetic database such that each of the five pathological classes is represented by 10 ECGs. The 50 measured ECGs were randomly selected from five 
           subsets of the PTB-XL database, 10 cases per subset, where each subset only contained signals labeled as 100\% pathological according to the 5 classes. Clinicians could choose from a list of 11 
        labels.  
           Clinicians were asked to make at least one annotation for each of the 100 pathological 12~lead ECG signals from a list of 11 pathologies as listed below:
           
           \begin{multicols}{2}
                \begin{itemize}
                    \item 1AVB
                    \item atrial fibrillation (AFIB)
                    \item FAM
                    \item IAB
                    \item LAO
                    \item LBBB
                    \item MI
                    \item normal healthy control (NORM)
                    \item right atrial overload/enlargement (RAO/RAE)
                    \item RBBB
                    \item Wolf-Parkinson-White syndrome (WPW)
                    \item[\vspace{\fill}]
                \end{itemize}
           \end{multicols}
           A total of two cardiologists responded. 

        \subsubsection*{Results}
            
\begin{figure}
    \centering
    \includegraphics{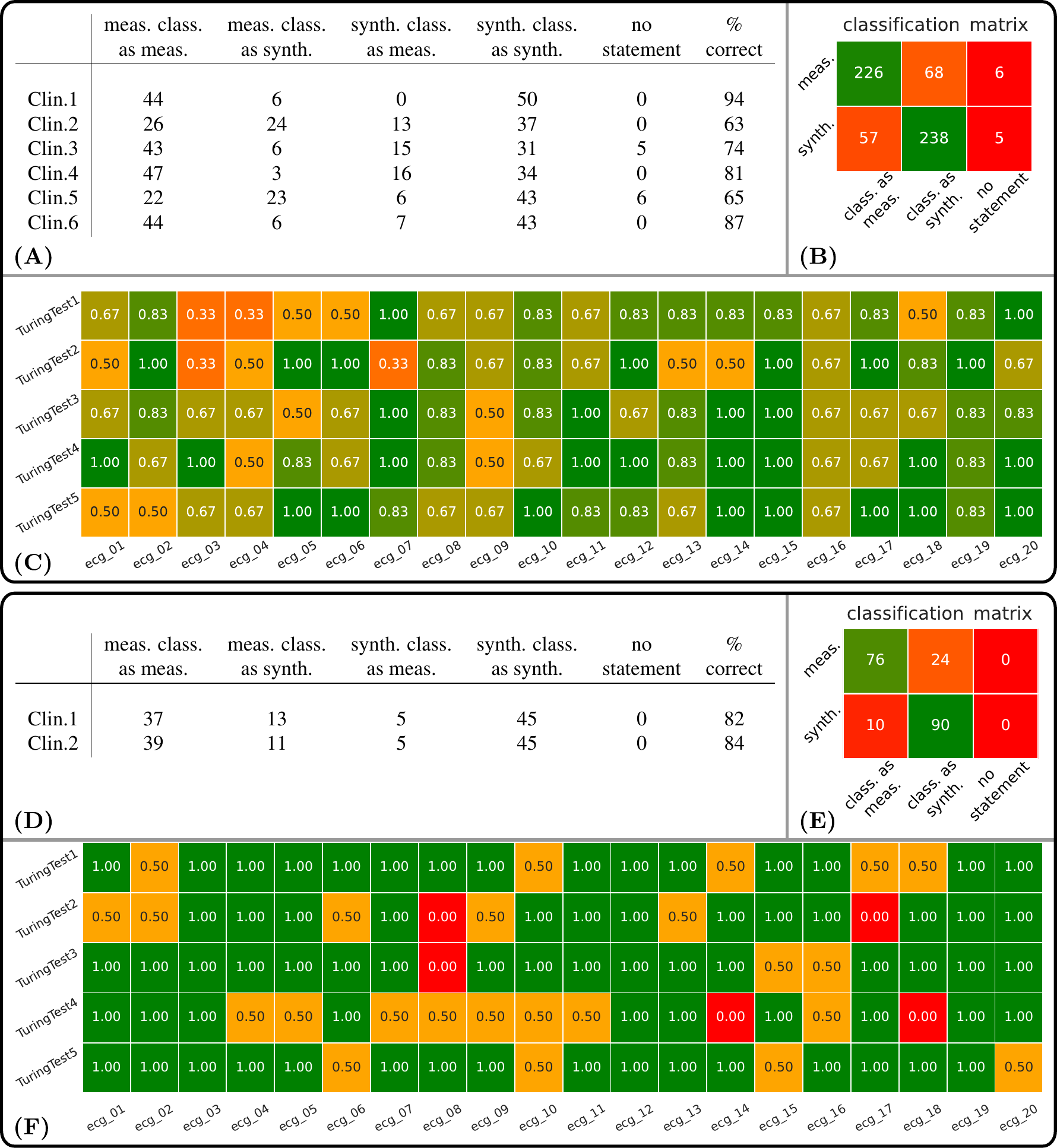}
    \caption{(Type classification) \textbf{Healthy cases}:
            \textbf{(A)} Classification results for each of the six expert 
             clinicians for the five Turing tests and percentage of correct 
             assessments. In summary, 62 of 300 assessments of the synthetic 
             ECGs and 74 of 300 assessments of the measured ECGs could not be 
             correctly classified by the experts. \textbf{(B)} Type classification 
             matrix across all 600 assessments. \textbf{(C)} Results of the clinical
             Turing tests performed by 6 clinicians. Each row corresponds to a 
             clinical Turing test and each square belongs to one of the 20 ECGs 
             per test. Shown is the relative number of clinicians who correctly 
             classified the corresponding signal.
              \textbf{Pathological cases}:
              \textbf{(D)} Type classification results for each of the two expert
             clinicians for the five Turing tests and percentage of correct
             assessments. In summary, 10 of 100 assessments of the synthetic
             ECGs and 24 of 100 assessments of the measured ECGs could not 
             be correctly classified by the experts. \textbf{(E)} Type
             classification matrix across all 100 assessments. \textbf{(F)} 
             Results of the clinical Turing tests 
             performed by 2 clinicians. Each row corresponds to a clinical 
             Turing test and each square belongs to one of the 20 ECGs per 
             test. Shown is the relative number of clinicians who correctly 
             classified the type of the corresponding signal. }
    \label{fig:turingTest_type_class}
\end{figure}

            \paragraph*{Normal Healthy Control Clinical Turing Test}
                The six clinicians correctly classified 464 of the 600 cases, which corresponds to an accuracy of $77.33\%$. On the other side, 136 signals ($22.67\%$) could not be correctly classified, including 62 ($10.34\%$) synthetic and 74 ($12.33\%$) measured 
                ECGs, see Figure~\ref{fig:turingTest_type_class}~{(B)}. 
                A detailed summary is
                given in Figure~\ref{fig:turingTest_type_class}~{(A),(C)}. Primary ECG features leading to classification as simulated included fractionation or improper R~wave propagation in the QRS~complex, a spiking or biphasic T~wave, and a lack of physiological noise in the signals. 
            
\begin{figure}
    \centering
    \includegraphics{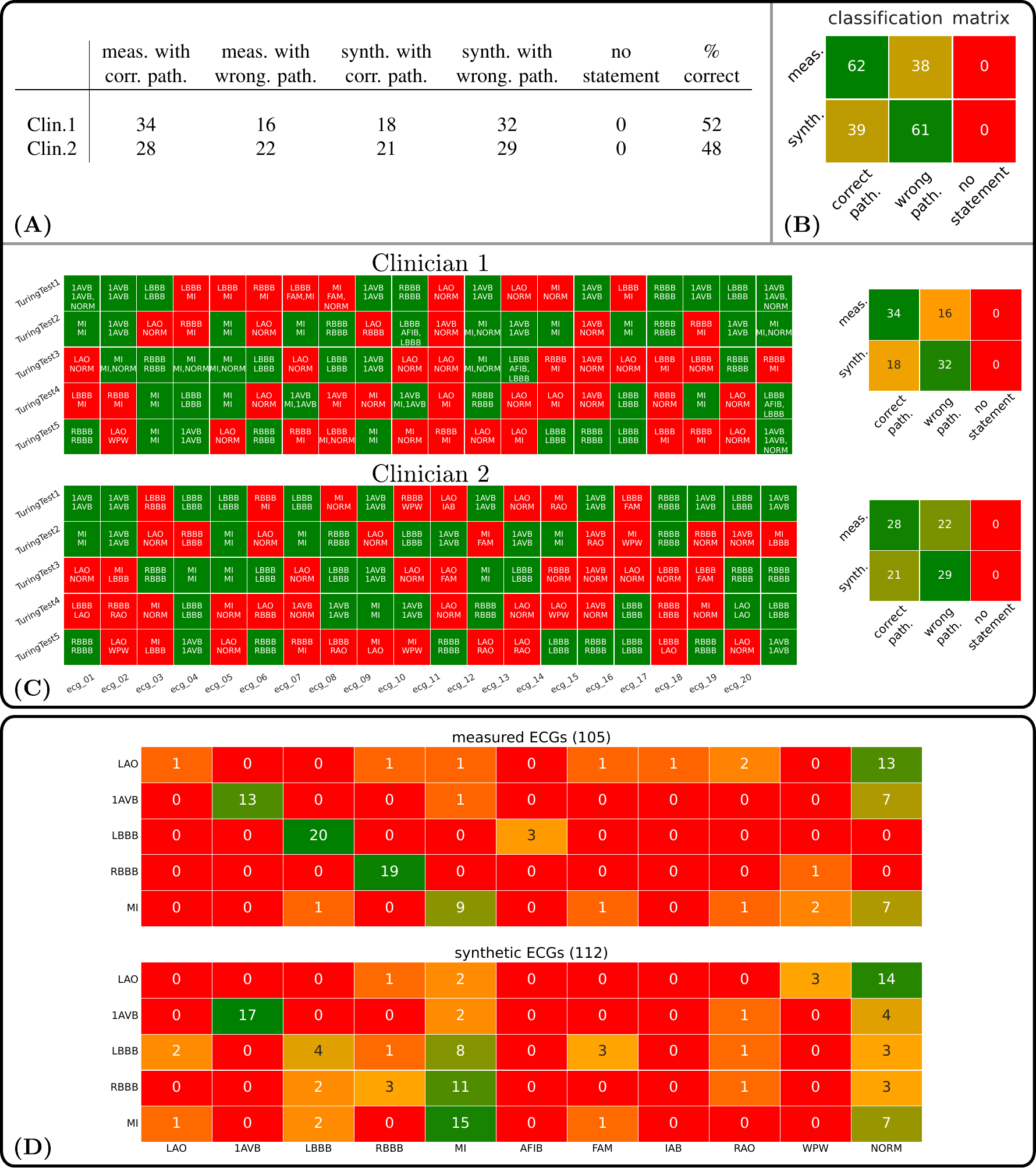}
    \caption{(Pathology classification) \textbf{(A)} Pathology classification
             results for each of the two expert clinicians for the five Turing 
             tests and percentage of correct assessments. In summary, 61 of 100
             assessments of the synthetic ECGs and 38 of 100 assessments of the 
             measured ECGs could not  be correctly classified by the experts.
             \textbf{(B)} Pathology classification matrix across all 100 assessments. 
             \textbf{(C)} (Clinician-based). Shown are the  classifications for both
             clinicians of all ECG Signals. For each ECG signal designated by a s
             quare, the top entries are the correct pathology and the bottom entries
             are the pathology actually selected by the user. Each row corresponds 
             to a clinical Turing test and each  square belongs to one of the 20 
             ECGs per test. \textbf{(D)} Confusion Matrices.}
    \label{fig:turingTest_path_class}
\end{figure}

            \paragraph*{Turing Test of Pathological ECGs}
                The two clinicians correctly classified the signals as either measured or clinical in 166 of the 200 cases, which 
                corresponds to an overall accuracy of $83\%$. On the other side, the type of 34 signals ($17\%$) could 
                not be correctly classified, including 10 ($5\%$) synthetic and 24 ($12\%$) measured 
                ECGs, see Figure~\ref{fig:turingTest_type_class}~{(E)}. A detailed summary is
                given in Figure~\ref{fig:turingTest_type_class}~{(D),(F)}. Regarding the correct classification of pathological cases, only 101 of the 200 ($50.5\%$) overall cases including both simulated and clinical signals were classified correctly by both clinicians. Namely, 
                38 measured ECGs were classified as the wrong pathology by experts resulting in an accuracy of $62\%$. Inversely, simulated pathologies were correctly classified at only $39\%$, with 61 signals being classified incorrectly. A detailed 
                summary is given in Figure~\ref{fig:turingTest_path_class}~{(A),(B)}. 
        
                The actual pathology and the diagnoses given by each clinician within the pathological clinical Turing test is provided in Figure~\ref{fig:turingTest_path_class}~(C). 
                Some pathologies were more commonly misdiagnosed by the clinicians and mistaken as either normal healthy control or an alternative clinical pathology. Differences in performance were also observed between simulated and clinical ECG sets. This is highlighted by the confusion matrices constructed for all pathological cases from the results for both measured and simulated signals (Figure~\ref{fig:turingTest_path_class}~(D)). 
                    
                Within clinical signals, the pathological cases of LAO, 1AVB, and MI were commonly mistaken as a 12~lead ECG in normal sinus rhythm by both clinicians. 
                Largest differences in diagnostic outcomes between simulated and clinical data sets is observed for LBBB and RBBB.
                Within simulated ECGs, LBBB and RBBB were commonly mistaken for MI.

\subsection*{Limitations and Summary}
    
    The feature analysis showed that the synthetic signals exhibit interval and amplitude features that are mostly in line with feature ranges reported in PTB-XL for the healthy and the pathological cohorts. However, they neither cover the full range of feature values that occur in clinical practice nor are they characterized by accurately coinciding distributions. 
    This could be attributed to the fact that the atrial model population was parameterized using ECG biomarker ranges for P~wave amplitudes and durations reported for extensive clinical cohorts partially comprising $>$200,000 subjects~\cite{nielsen15,unknown-0000-ID14904} which might lead to slightly different feature distributions compared to those extractable from PTB-XL. The QRST~complexes were also parameterized according to experimental data or clinical data conducted on smaller model cohorts that may not be representative of the entire population. Some parameters were also estimated as no direct clinical or experimental data is available for these entities. 
    One such example is the heightened T~wave amplitudes, which stem from repolarization gradients in the ventricles that generate large cardiac source. While the occurrence of repolarization gradients are known \cite{opthof2017cardiac,opthof2016dispersion}, the exact nature of such gradients are not well understood and thus hard to parameterize for a patient population. 
    Therefore, the synthetic signals are not fully representative for an entire population, such as the one in PTB-XL. 
    
    The feature distributions in the synthetic cohort are however consistent in themselves, i.e., unrealistic combinations of different features are unlikely to occur. For example, the upper limit of RR~intervals in the simulated healthy cohort does not exceed 1000\,$\si{\milli \second}$, while simultaneously, the QT~interval also only covers lower ranges of the clinical QT~interval values (compare Figure~\ref{fig:comparisonFeaturesHealthy}). This is due to the fact that multi-variate normal distributions were used during the synthesization procedure ensuring that clinically reported correlations between ECG biomarkers (such as P~wave duration and PR~interval or QT~duration and RR~intervals) are taken into account. Furthermore, detailed mechanistic electrophysiological models of the heart were employed and simulation parameters in reasonable ranges reported in literature were chosen leading to realistic single beat P~waves and QRST~complexes in most cases. 
    
\begin{landscape}
\begin{figure}[ht]
\centering
\includegraphics[width=\linewidth]{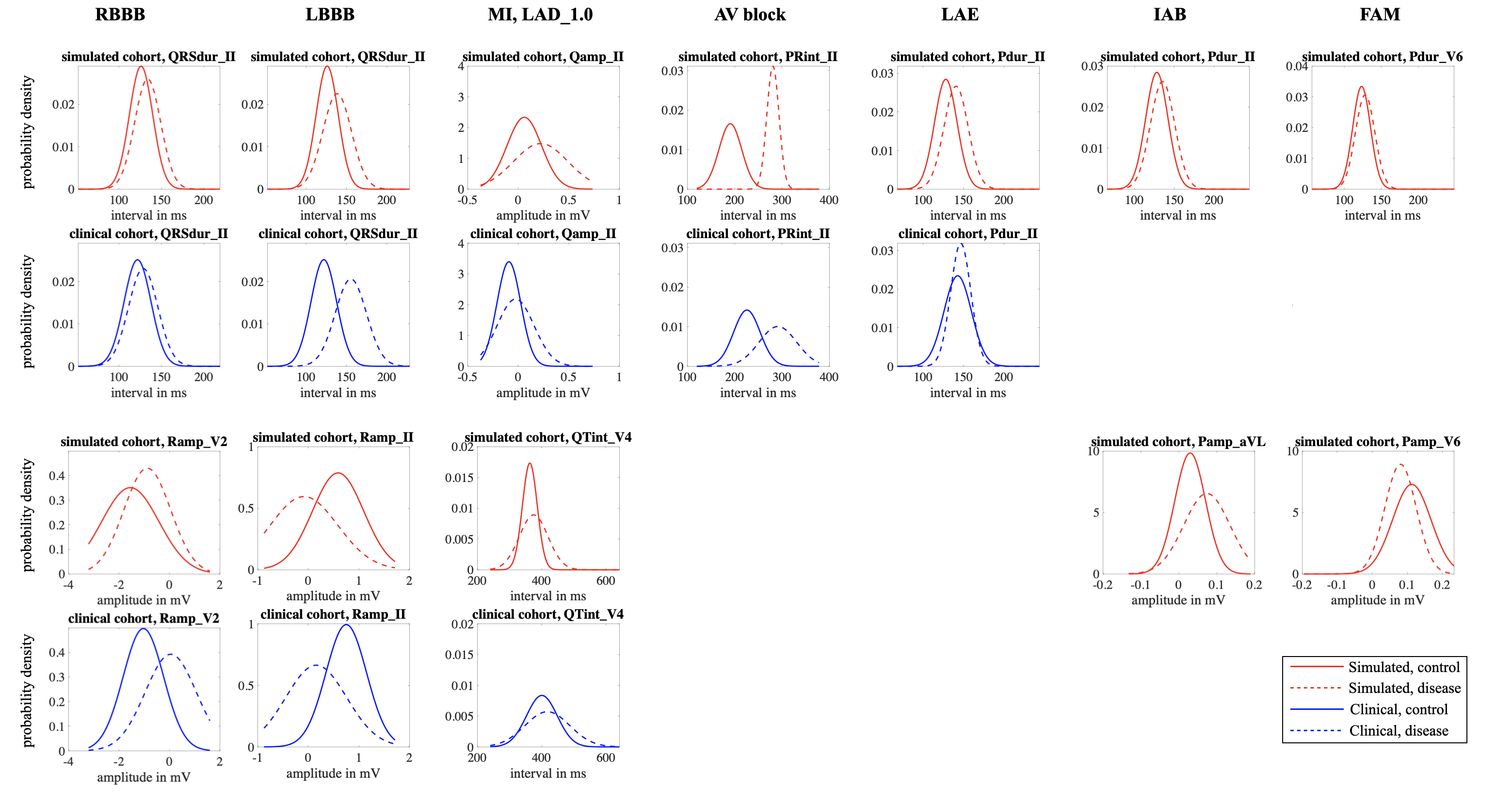}
\caption{Comparison of features extracted from healthy (solid lines) and pathological (dotted line) ECGs in the clinical (blue curves, bottom panel) and virtual (red curve, top panel) cohorts. Probability density functions are shown for selected timing or amplitude features that are clinically evaluated for a diagnosis of the displayed disease (from left to right: RBBB, LBBB, MI, 1AVB, LAO, IAB and FAM).}
\label{fig:comparisonFeaturesDiseases}
\end{figure}
\end{landscape}

    It must be noted that PTB-XL lacks clinical data for fibrotic atrial cardiomyopathy and for interatrial conduction block. Thus, fidelity assessment of ECG features within these two classes by means of a comparison to clinical data was not possible using the same clinical ECG resources. 
    However, we already showed in previous work that the simulated P~waves reproduce characteristic changes in key diagnostic ECG markers~\cite{Nagel-2021-ID16114, Bender-2022-ID18178}. These include a prolongation of the P~wave duration compared to the control simulations due to delayed depolarization in fibrotic patches as well as a retrograde activation of the left atrium through interatrial conduction pathways on the posterior wall. Moreover, as shown in Figure~\ref{fig:comparisonFeaturesDiseases}, in interatrial conduction block patients, the morphology and therefore the P~wave amplitude is markedly changed in lead aVL compared to the healthy cohort. In patients with fibrotic atrial cardiomyopathy, the most pronounced decrease in P~wave amplitude due to scar tissue not contributing to the overall source distribution in the atria occurs in the lateral leads (compare Figure~\ref{fig:comparisonFeaturesDiseases}). 
      
    The clinical Turing tests aimed to investigate the ability of the 12~lead ECG 
    signal to exhibit morphological features in accordance with clinical diagnostic 
    criteria as routinely assessed by clinicians under both normal healthy control and pathological conditions. Within the clinical Turing test performed for normal healthy control, it can be observed that accuracy in identifying whether a signal was simulated or clinical was 77$\%$ accurate. Primary ECG features leading to identification as a synthetic signal included fractionation and R~wave progression of the QRS~complex. Spiked T~waves with high amplitudes or biphasic T~waves could also be observed. Real ECG signals tended to also exhibit a certain noise types not accounted for, including electrical disturbances and large baseline wander, that must either be modulated within simulated data or removed during the clinical Turing test.
    Within the clinical Turing test to diagnose pathological ECGs, the accuracy of type classification increased to 83$\%$, indicating type classification was easier with synthetic pathological data. Misdiagnosis was common across both signal types as pathologies were only diagnosed correctly by the two expert cardiologists in $51\%$ of cases. More clinicians should perform the clinical Turing test on pathology classification to give a better indication of the true accuracy of ECG diagnosis on both simulated and clinical signals. Furthermore, the clinical Turing test must be conducted on a larger number signals beyond the 100 analyzed, ideally,  for the entire ECG synthetic database. 
    
    Regardless, it can be observed that clinicians had varying performance on clinical-based 12~lead ECG signals in comparison to those taken from the synthetic ECG database. Clinical signals were classified with the correct pathology at an accuracy of $62\%$. Simulated signals, on the other side, were classified correctly for the underlying disease pathology at only $39\%$. LAO within both clinical and simulated data experienced the highest level of misdiagnosis and resulted in similar performance. This could be contributed to the fact that LAO manifests only within the P~wave, where morphological deviations are harder to detect due to a substantially lower amplitude than the QRS complex. Misdiagnosis was also high among the diseases of LBBB and RBBB within the simulated data set. Differences in outcome between the clinical and synthetic signals may stem from the inability of the synthetic ECG database to manifest the full complexity of the underlying diseases. For example, remodeling within the ventricles under such conditions may lead to slower conduction properties and alternative wave morphology. Furthermore, only complete LBBB or RBBB was modeled. In clinical practice, however, there are varying degrees of conduction block. A lower reported diagnostic accuracy for MI and 1AVB is seen for the clinical signals in comparison to the simulated ECGs, which could also stem from a lack of complexity within the simulated setup easing diagnosis. 
    
    To lower the mismatch in performance between clinical and synthetic signals, further parameter tuning is needed. Iterative clinical Turing tests would be beneficial to update parameters ranges to mitigate the prevalence of undesirable ECG features within the entire database. Refinement could also be guided by sensitivity analysis that provides more information on the relationship of model parameters and the morphological traits of simulated signals as determined by clinicians.  However, this requires a large investment due to the variety in clinical pathological classes, and the lack of known electrophysiology in such conditions. Certain important ECG features may also be  detected  by machine learning analysis \cite{strodthoff2020deep} to provide insight into the  refined  sub-classification of  pathological cases beyond current routine diagnoses. 
    
    Some results from the Turing test of pathological cases indicate that standard protocols for ECG classification by clinicians are not sufficient. Machine learning algorithms may offer a means to aide in ECG diagnosis to improve reliability of clinical decisions. Therefore it is important to provide reference data to test such algorithms.  An earlier benchmark study demonstrated this with the large data set of clinical ECGs in PTB-XL \cite{strodthoff2020deep}. In this work, deep learning algorithms were e. g. found to exhibit diagnosis success rates in the range of 80 – 95 percent depending on the used metric. The clinical PTB-XL data set was also instrumental in demonstrating the clear   improvement of algorithms based on self-supervised learning \cite{mehari2022self}.  Nevertheless, clinical data bases strongly depend on the quality and the terminology used to label the ECG data. In addition large sets of publicly available clinical data sets are still rare and limited in number. Here is where benchmarking ML algorithm with validated simulated data sets can become an important tool in the development and benchmarking of new algorithm for ECG classification. Machine learning algorithms could then also be trained and tested on real and synthetic data in different combinations.  Data bases of simulated ECGs like the MedalCare-XL set presented in this paper provide also an important link of the growing knowledge developed in the cardiac modelling community and practical development of algorithm for data analysis.

\section*{Usage Notes}

    When using the synthetic ECGs as an input data source for machine learning applications, samples that were generated based on the same anatomical model should explicitly belong to only one of the training, testing or validation sets. As the main variation in morphology of the P~waves and QRST~complexes stem predominantly from anatomical differences in the model cohort~\cite{Dossel-2021-ID16522}, splitting the data in the described fashion thus helps to prevent overfitting to similar or almost identical samples that were already seen during training~\cite{Luongo-2020-ID14752}. 
    
    When applying the simulated data for extending or replacing small or imbalanced clinical datasets, the user is advised to refer to the signals with superimposed realistic ECG noise instead of the raw signal traces. In this way, the simulated signals exhibit characteristics due to noise interference that are also observable in clinical ECGs. Thus, possible domain gaps can be reduced eventually leading to an improved classification outcome on actual clinical data. 
    

\section*{Code availability}

    The anatomical model cohort of the atria is publicly available under the Creative Commons Licence CC-BY 4.0~\cite{nagel_claudia_2020_4309958}. 
    Code for solving the Eikonal equation and the forward problem of electrocardiography using the boundary element method as used for the atrial simulations is openly available (Stenroos~\etal~\cite{stenroos07}, Schuler~\etal~\cite{steffen_schuler_2022_7217554}). Python code for synthesizing single beat P~waves and QRST~complexes to a 10$\SI{}{\second}$ time series using multi-variate normal distributions for amplitude scaling and interval selection is publicly available~\cite{ecg-syn}.
    
    The cohort of ventricular-torso models are not publicly available due to constraints of the IRB approval and subject consent. The framework to simulate electrophysiology of the ventricular-torso model is also not available for public use in its entirety. However, all simulations can be carried out within the publicly available openCARP simulation framework~\cite{plank2021opencarp,openCARP_sw}, albeit less efficiently with higher compute costs. 
    

\bibliography{medalcare_d1_references}

\begin{thebibliography}{10}
\urlstyle{rm}
\expandafter\ifx\csname url\endcsname\relax
  \def\url#1{\texttt{#1}}\fi
\expandafter\ifx\csname urlprefix\endcsname\relax\def\urlprefix{URL }\fi
\expandafter\ifx\csname doiprefix\endcsname\relax\def\doiprefix{DOI: }\fi
\providecommand{\bibinfo}[2]{#2}
\providecommand{\eprint}[2][]{\url{#2}}

\bibitem{wagner2020ptb}
\bibinfo{author}{Wagner, P.} \emph{et~al.}
\newblock \bibinfo{journal}{\bibinfo{title}{{PTB-XL}, a large publicly
  available electrocardiography dataset}}.
\newblock {\emph{\JournalTitle{Scientific Data}}} \textbf{\bibinfo{volume}{7}},
  \bibinfo{pages}{1--15} (\bibinfo{year}{2020}).

\bibitem{Roberts-2021-ID15918}
\bibinfo{author}{{R}oberts, M.} \emph{et~al.}
\newblock \bibinfo{journal}{\bibinfo{title}{{C}ommon pitfalls and
  recommendations for using machine learning to detect and prognosticate for
  {COVID}-19 using chest radiographs and {CT} scans}}.
\newblock {\emph{\JournalTitle{{N}ature {M}achine {I}ntelligence}}}
  \textbf{\bibinfo{volume}{3}}, \bibinfo{pages}{199--217},
  \url{10.1038/s42256-021-00307-0} (\bibinfo{year}{2021}).

\bibitem{Puyol-Anton-2021-ID17011}
\bibinfo{author}{Puyol-Ant{\'o}n, E.} \emph{et~al.}
\newblock \bibinfo{title}{Fairness in cardiac mr image analysis: An
  investigation of bias due to data imbalance in deep learning based
  segmentation}.
\newblock In \bibinfo{editor}{de~Bruijne, M.} \emph{et~al.} (eds.)
  \emph{\bibinfo{booktitle}{Medical Image Computing and Computer Assisted
  Intervention -- MICCAI 2021}}, \bibinfo{pages}{413--423}
  (\bibinfo{publisher}{Springer International Publishing},
  \bibinfo{address}{Cham}, \bibinfo{year}{2021}).

\bibitem{Pilia-2020-ID14680}
\bibinfo{author}{{P}ilia, N.} \emph{et~al.}
\newblock \bibinfo{journal}{\bibinfo{title}{{Q}uantification and classification
  of potassium and calcium disorders with the electrocardiogram: {W}hat do
  clinical studies, modeling, and reconstruction tell us?}}
\newblock {\emph{\JournalTitle{{APL} {B}ioeng}}} \textbf{\bibinfo{volume}{4}},
  \bibinfo{pages}{041501}, \url{10.1063/5.0018504} (\bibinfo{year}{2020}).

\bibitem{Luongo-2022-ID17367}
\bibinfo{author}{{L}uongo, G.} \emph{et~al.}
\newblock \bibinfo{journal}{\bibinfo{title}{{H}ybrid machine learning to
  localize atrial flutter substrates using the surface 12-lead
  electrocardiogram}}.
\newblock {\emph{\JournalTitle{{EP} {Europace}}}}
  \url{10.1093/europace/euab322} (\bibinfo{year}{2022}).

\bibitem{Nagel-2022-ID17346}
\bibinfo{author}{Nagel, C.}, \bibinfo{author}{Schaufelberger, M.},
  \bibinfo{author}{D{\"o}ssel, O.} \& \bibinfo{author}{Loewe, A.}
\newblock \bibinfo{title}{A bi-atrial statistical shape model as a basis
  to classify left atrial enlargement from simulated and clinical 12-lead
  {ECG}s}.
\newblock In \bibinfo{editor}{Puyol~Ant{\'o}n, E.} \emph{et~al.} (eds.)
  \emph{\bibinfo{booktitle}{Statistical Atlases and Computational Models of the
  Heart. Multi-Disease, Multi-View, and Multi-Center Right Ventricular
  Segmentation in Cardiac MRI Challenge}}, \bibinfo{pages}{38--47},
  \url{10.1007/978-3-030-93722-5_5} (\bibinfo{year}{2022}).

\bibitem{Luongo-2021-ID15954}
\bibinfo{author}{{L}uongo, G.} \emph{et~al.}
\newblock \bibinfo{journal}{\bibinfo{title}{{M}achine learning enables
  noninvasive prediction of atrial fibrillation driver location and acute
  pulmonary vein ablation success using the 12-lead {ECG}}}.
\newblock {\emph{\JournalTitle{{C}ardiovascular {D}igital {H}ealth {J}ournal}}}
  \textbf{\bibinfo{volume}{2}}, \bibinfo{pages}{126--136},
  \url{10.1016/j.cvdhj.2021.03.002} (\bibinfo{year}{2021}).

\bibitem{strocchi2020publicly}
\bibinfo{author}{Strocchi, M.} \emph{et~al.}
\newblock \bibinfo{journal}{\bibinfo{title}{A publicly available virtual cohort
  of four-chamber heart meshes for cardiac electro-mechanics simulations}}.
\newblock {\emph{\JournalTitle{PloS one}}} \textbf{\bibinfo{volume}{15}},
  \bibinfo{pages}{e0235145} (\bibinfo{year}{2020}).

\bibitem{american2002standardized}
\bibinfo{author}{{American Heart Association Writing Group on Myocardial
  Segmentation and Registration for Cardiac Imaging}} \emph{et~al.}
\newblock \bibinfo{journal}{\bibinfo{title}{Standardized myocardial
  segmentation and nomenclature for tomographic imaging of the heart: statement
  for healthcare professionals from the cardiac imaging committee of the
  council on clinical cardiology of the american heart association}}.
\newblock {\emph{\JournalTitle{Circulation}}} \textbf{\bibinfo{volume}{105}},
  \bibinfo{pages}{539--542} (\bibinfo{year}{2002}).

\bibitem{medalcare-xl}
\bibinfo{author}{Gillette, K.} \emph{et~al.}
\newblock \bibinfo{title}{{MedalCare-XL}}, \url{10.5281/zenodo.7293655}
  (\bibinfo{year}{2022}).

\bibitem{SCI:Seg3D}
\bibinfo{author}{CIBC} (\bibinfo{year}{2016}).
\newblock \bibinfo{note}{Seg3D: Volumetric image segmentation and
  visualization. Scientific Computing and Imaging http://www.seg3d.org}.

\bibitem{payer2017multi}
\bibinfo{author}{Payer, C.}, \bibinfo{author}{{\v{S}}tern, D.},
  \bibinfo{author}{Bischof, H.} \& \bibinfo{author}{Urschler, M.}
\newblock \bibinfo{title}{Multi-label whole heart segmentation using cnns and
  anatomical label configurations}.
\newblock In \emph{\bibinfo{booktitle}{International Workshop on Statistical
  Atlases and Computational Models of the Heart}}, \bibinfo{pages}{190--198}
  (\bibinfo{organization}{Springer}, \bibinfo{year}{2017}).

\bibitem{chetverikov2002trimmed}
\bibinfo{author}{Chetverikov, D.}, \bibinfo{author}{Svirko, D.},
  \bibinfo{author}{Stepanov, D.} \& \bibinfo{author}{Krsek, P.}
\newblock \bibinfo{title}{The trimmed iterative closest point algorithm}.
\newblock In \emph{\bibinfo{booktitle}{Pattern Recognition, 2002. Proceedings.
  16th International Conference on}}, vol.~\bibinfo{volume}{3},
  \bibinfo{pages}{545--548} (\bibinfo{organization}{IEEE},
  \bibinfo{year}{2002}).

\bibitem{prassl2009automatically}
\bibinfo{author}{Prassl, A.~J.} \emph{et~al.}
\newblock \bibinfo{journal}{\bibinfo{title}{Automatically generated,
  anatomically accurate meshes for cardiac electrophysiology problems}}.
\newblock {\emph{\JournalTitle{IEEE Transactions on Biomedical Engineering}}}
  \textbf{\bibinfo{volume}{56}}, \bibinfo{pages}{1318--1330}
  (\bibinfo{year}{2009}).

\bibitem{gillette2021framework}
\bibinfo{author}{Gillette, K.} \emph{et~al.}
\newblock \bibinfo{journal}{\bibinfo{title}{A framework for the generation of
  digital twins of cardiac electrophysiology from clinical 12-leads {ECG}s}}.
\newblock {\emph{\JournalTitle{Medical Image Analysis}}}
  \textbf{\bibinfo{volume}{71}}, \bibinfo{pages}{102080}
  (\bibinfo{year}{2021}).

\bibitem{bayer2018universal}
\bibinfo{author}{Bayer, J.} \emph{et~al.}
\newblock \bibinfo{journal}{\bibinfo{title}{Universal ventricular coordinates:
  A generic framework for describing position within the heart and transferring
  data}}.
\newblock {\emph{\JournalTitle{Medical Image Analysis}}}
  \textbf{\bibinfo{volume}{45}}, \bibinfo{pages}{83--93}
  (\bibinfo{year}{2018}).

\bibitem{Nagel-2021-ID16581}
\bibinfo{author}{{N}agel, C.}, \bibinfo{author}{{S}chuler, S.},
  \bibinfo{author}{{D}össel, O.} \& \bibinfo{author}{{L}oewe, A.}
\newblock \bibinfo{journal}{\bibinfo{title}{{A} bi-atrial statistical shape
  model for large-scale in silico studies of human atria: Model development and
  application to {ECG} simulations}}.
\newblock {\emph{\JournalTitle{{M}edical {I}mage {A}nalysis}}}
  \textbf{\bibinfo{volume}{74}}, \bibinfo{pages}{102210},
  \url{10.1016/j.media.2021.102210} (\bibinfo{year}{2021}).

\bibitem{nagel_claudia_2020_4309958}
\bibinfo{author}{{N}agel, C.}, \bibinfo{author}{{S}chuler, S.},
  \bibinfo{author}{{D}össel, O.} \& \bibinfo{author}{{L}oewe, A.}
\newblock \bibinfo{journal}{\bibinfo{title}{{A} bi-atrial statistical shape
  model and 100 volumetric anatomical models of the atria}}.
\newblock {\emph{\JournalTitle{Zenodo}}} \url{10.5281/zenodo.4309958}
  (\bibinfo{year}{2020}).

\bibitem{Azzolin-2022-ID17490}
\bibinfo{author}{{A}zzolin, L.} \emph{et~al.}
\newblock \bibinfo{journal}{\bibinfo{title}{{A}ugment{A}: Patient-specific
  augmented atrial model generation tool}}.
\newblock {\emph{\JournalTitle{med{R}xiv}}} \url{10.1101/2022.02.13.22270835}
  (\bibinfo{year}{2022}).

\bibitem{Zheng-2021-ID17989}
\bibinfo{author}{{Z}heng, T.}, \bibinfo{author}{{A}zzolin, L.},
  \bibinfo{author}{{S}ánchez, J.}, \bibinfo{author}{{D}össel, O.} \&
  \bibinfo{author}{{L}oewe, A.}
\newblock \bibinfo{journal}{\bibinfo{title}{{A}n automate pipeline for
  generating fiber orientation and region annotation in patient specific atrial
  models}}.
\newblock {\emph{\JournalTitle{{C}urrent {D}irections in {B}iomedical
  {E}ngineering}}} \textbf{\bibinfo{volume}{7}}, \bibinfo{pages}{136--139},
  \url{10.1515/cdbme-2021-2035} (\bibinfo{year}{2021}).

\bibitem{Lang-2015-ID16479}
\bibinfo{author}{{L}ang, R.~M.} \emph{et~al.}
\newblock \bibinfo{journal}{\bibinfo{title}{{R}ecommendations for cardiac
  chamber quantification by echocardiography in adults: An update from the
  {A}merican {S}ociety of {E}chocardiography and the {E}uropean {A}ssociation
  of {C}ardiovascular {I}maging.}}
\newblock {\emph{\JournalTitle{{E}ur {H}eart {J} {C}ardiovasc {I}maging}}}
  \textbf{\bibinfo{volume}{16}}, \bibinfo{pages}{233--70},
  \url{10.1093/ehjci/jev014} (\bibinfo{year}{2015}).

\bibitem{Nagel-2021-ID16114}
\bibinfo{author}{{N}agel, C.} \emph{et~al.}
\newblock \bibinfo{journal}{\bibinfo{title}{{N}on-invasive and quantitative
  estimation of left atrial fibrosis based on {P} waves of the 12-lead {ECG} -
  a large-scale computational study covering anatomical variability.}}
\newblock {\emph{\JournalTitle{{J} {C}lin {M}ed}}}
  \textbf{\bibinfo{volume}{10}}, \url{10.3390/jcm10081797}
  (\bibinfo{year}{2021}).

\bibitem{Pishchulin-2017-ID12226}
\bibinfo{author}{{P}ishchulin, L.}, \bibinfo{author}{{W}uhrer, S.},
  \bibinfo{author}{{H}elten, T.}, \bibinfo{author}{{T}heobalt, C.} \&
  \bibinfo{author}{{S}chiele, B.}
\newblock \bibinfo{journal}{\bibinfo{title}{{B}uilding statistical shape spaces
  for 3{D} human modeling}}.
\newblock {\emph{\JournalTitle{{P}attern {R}ecognition}}}
  \textbf{\bibinfo{volume}{67}}, \bibinfo{pages}{276--286},
  \url{10.1016/j.patcog.2017.02.018} (\bibinfo{year}{2017}).

\bibitem{durrer1970total}
\bibinfo{author}{Durrer, D.} \emph{et~al.}
\newblock \bibinfo{journal}{\bibinfo{title}{Total excitation of the isolated
  human heart}}.
\newblock {\emph{\JournalTitle{Circulation}}} \textbf{\bibinfo{volume}{41}},
  \bibinfo{pages}{899--912} (\bibinfo{year}{1970}).

\bibitem{kassebaum1966electrophysiological}
\bibinfo{author}{Kassebaum, D.~G.} \& \bibinfo{author}{Van~Dyke, A.~R.}
\newblock \bibinfo{journal}{\bibinfo{title}{Electrophysiological effects of
  isoproterenol on purkinje fibers of the heart}}.
\newblock {\emph{\JournalTitle{Circulation Research}}}
  \textbf{\bibinfo{volume}{19}}, \bibinfo{pages}{940--946}
  (\bibinfo{year}{1966}).

\bibitem{bayer2012novel}
\bibinfo{author}{Bayer, J.~D.}, \bibinfo{author}{Blake, R.~C.},
  \bibinfo{author}{Plank, G.} \& \bibinfo{author}{Trayanova, N.~A.}
\newblock \bibinfo{journal}{\bibinfo{title}{A novel rule-based algorithm for
  assigning myocardial fiber orientation to computational heart models}}.
\newblock {\emph{\JournalTitle{Annals of Biomedical Engineering}}}
  \textbf{\bibinfo{volume}{40}}, \bibinfo{pages}{2243--2254}
  (\bibinfo{year}{2012}).

\bibitem{streeter1969fiber}
\bibinfo{author}{Streeter~Jr, D.~D.}, \bibinfo{author}{Spotnitz, H.~M.},
  \bibinfo{author}{Patel, D.~P.}, \bibinfo{author}{Ross~Jr, J.} \&
  \bibinfo{author}{Sonnenblick, E.~H.}
\newblock \bibinfo{journal}{\bibinfo{title}{Fiber orientation in the canine
  left ventricle during diastole and systole}}.
\newblock {\emph{\JournalTitle{Circulation Research}}}
  \textbf{\bibinfo{volume}{24}}, \bibinfo{pages}{339--347}
  (\bibinfo{year}{1969}).

\bibitem{taggart2000inhomogeneous}
\bibinfo{author}{Taggart, P.} \emph{et~al.}
\newblock \bibinfo{journal}{\bibinfo{title}{Inhomogeneous transmural conduction
  during early ischaemia in patients with coronary artery disease}}.
\newblock {\emph{\JournalTitle{Journal of Molecular and Cellular Cardiology}}}
  \textbf{\bibinfo{volume}{32}}, \bibinfo{pages}{621--630}
  (\bibinfo{year}{2000}).

\bibitem{roberts1982effect}
\bibinfo{author}{Roberts, D.~E.} \& \bibinfo{author}{Scher, A.~M.}
\newblock \bibinfo{journal}{\bibinfo{title}{Effect of tissue anisotropy on
  extracellular potential fields in canine myocardium in situ.}}
\newblock {\emph{\JournalTitle{Circulation Research}}}
  \textbf{\bibinfo{volume}{50}}, \bibinfo{pages}{342--351}
  (\bibinfo{year}{1982}).

\bibitem{keller2010ranking}
\bibinfo{author}{Keller, D.~U.}, \bibinfo{author}{Weber, F.~M.},
  \bibinfo{author}{Seemann, G.} \& \bibinfo{author}{D{\"o}ssel, O.}
\newblock \bibinfo{journal}{\bibinfo{title}{Ranking the influence of tissue
  conductivities on forward-calculated {ECG}s}}.
\newblock {\emph{\JournalTitle{IEEE Transactions on Biomedical Engineering}}}
  \textbf{\bibinfo{volume}{57}}, \bibinfo{pages}{1568--1576}
  (\bibinfo{year}{2010}).

\bibitem{mitchell2003two}
\bibinfo{author}{Mitchell, C.~C.} \& \bibinfo{author}{Schaeffer, D.~G.}
\newblock \bibinfo{journal}{\bibinfo{title}{A two-current model for the
  dynamics of cardiac membrane}}.
\newblock {\emph{\JournalTitle{Bulletin of Mathematical Biology}}}
  \textbf{\bibinfo{volume}{65}}, \bibinfo{pages}{767--793}
  (\bibinfo{year}{2003}).

\bibitem{opthof2017cardiac}
\bibinfo{author}{Opthof, T.} \emph{et~al.}
\newblock \bibinfo{journal}{\bibinfo{title}{Cardiac activation--repolarization
  patterns and ion channel expression mapping in intact isolated normal human
  hearts}}.
\newblock {\emph{\JournalTitle{Heart Rhythm}}} \textbf{\bibinfo{volume}{14}},
  \bibinfo{pages}{265--272} (\bibinfo{year}{2017}).

\bibitem{opthof2016dispersion}
\bibinfo{author}{Opthof, T.} \emph{et~al.}
\newblock \bibinfo{journal}{\bibinfo{title}{Dispersion in ventricular
  repolarization in the human, canine and porcine heart}}.
\newblock {\emph{\JournalTitle{Progress in Biophysics and Molecular Biology}}}
  \textbf{\bibinfo{volume}{120}}, \bibinfo{pages}{222--235}
  (\bibinfo{year}{2016}).

\bibitem{keller2011influence}
\bibinfo{author}{Keller, D.~U.}, \bibinfo{author}{Weiss, D.~L.},
  \bibinfo{author}{Dossel, O.} \& \bibinfo{author}{Seemann, G.}
\newblock \bibinfo{journal}{\bibinfo{title}{Influence of {$I_{Ks}$}
  heterogeneities on the genesis of the t-wave: A computational evaluation}}.
\newblock {\emph{\JournalTitle{IEEE Transactions on Biomedical Engineering}}}
  \textbf{\bibinfo{volume}{59}}, \bibinfo{pages}{311--322}
  (\bibinfo{year}{2011}).

\bibitem{neic2020meshtool}
\bibinfo{author}{Neic, A.}, \bibinfo{author}{Gsell, M. A.~F.},
  \bibinfo{author}{Karabelas, E.}, \bibinfo{author}{Prassl, A.~J.} \&
  \bibinfo{author}{Plank, G.}
\newblock \bibinfo{journal}{\bibinfo{title}{{Automating image-based mesh
  generation and manipulation tasks in cardiac modeling workflows using
  Meshtool.}}}
\newblock {\emph{\JournalTitle{SoftwareX}}} \textbf{\bibinfo{volume}{11}},
  \bibinfo{pages}{100454}, \url{10.1016/j.softx.2020.100454}
  (\bibinfo{year}{2020}).

\bibitem{mendonca2018modeling}
\bibinfo{author}{Mendonca~Costa, C.}, \bibinfo{author}{Plank, G.},
  \bibinfo{author}{Rinaldi, C.~A.}, \bibinfo{author}{Niederer, S.~A.} \&
  \bibinfo{author}{Bishop, M.~J.}
\newblock \bibinfo{journal}{\bibinfo{title}{Modeling the electrophysiological
  properties of the infarct border zone}}.
\newblock {\emph{\JournalTitle{Frontiers in Physiology}}}
  \textbf{\bibinfo{volume}{9}}, \bibinfo{pages}{356} (\bibinfo{year}{2018}).

\bibitem{loewe18}
\bibinfo{author}{{L}oewe, A.}, \bibinfo{author}{{W}ülfers, E.~M.} \&
  \bibinfo{author}{{S}eemann, G.}
\newblock \bibinfo{journal}{\bibinfo{title}{{C}ardiac ischemia-insights from
  computational models.}}
\newblock {\emph{\JournalTitle{{H}erzschrittmacher \& {E}lektrophysiologie}}}
  \textbf{\bibinfo{volume}{29}}, \bibinfo{pages}{48--56},
  \url{10.1007/s00399-017-0539-6} (\bibinfo{year}{2018}).

\bibitem{neic2017reaction_eikonal}
\bibinfo{author}{Neic, A.} \emph{et~al.}
\newblock \bibinfo{journal}{\bibinfo{title}{Efficient computation of
  electrograms and {ECG}s in human whole heart simulations using a
  reaction-eikonal model.}}
\newblock {\emph{\JournalTitle{Journal of Computational Physics}}}
  \textbf{\bibinfo{volume}{346}}, \bibinfo{pages}{191--211},
  \url{10.1016/j.jcp.2017.06.020} (\bibinfo{year}{2017}).

\bibitem{potse2018scalable}
\bibinfo{author}{Potse, M.}
\newblock \bibinfo{journal}{\bibinfo{title}{Scalable and accurate {ECG}
  simulation for reaction-diffusion models of the human heart}}.
\newblock {\emph{\JournalTitle{Frontiers in Physiology}}}
  \textbf{\bibinfo{volume}{9}}, \bibinfo{pages}{370} (\bibinfo{year}{2018}).

\bibitem{vigmond2008solvers}
\bibinfo{author}{Vigmond, E.}, \bibinfo{author}{Dos~Santos, R.~W.},
  \bibinfo{author}{Prassl, A.}, \bibinfo{author}{Deo, M.} \&
  \bibinfo{author}{Plank, G.}
\newblock \bibinfo{journal}{\bibinfo{title}{Solvers for the cardiac bidomain
  equations}}.
\newblock {\emph{\JournalTitle{Progress in Biophysics and Molecular Biology}}}
  \textbf{\bibinfo{volume}{96}}, \bibinfo{pages}{3--18} (\bibinfo{year}{2008}).

\bibitem{plank2021opencarp}
\bibinfo{author}{Plank, G.} \emph{et~al.}
\newblock \bibinfo{journal}{\bibinfo{title}{The {openCARP} simulation
  environment for cardiac electrophysiology}}.
\newblock {\emph{\JournalTitle{Computer Methods and Programs in Biomedicine}}}
  \textbf{\bibinfo{volume}{208}}, \bibinfo{pages}{106223},
  \url{10.1016/j.cmpb.2021.106223} (\bibinfo{year}{2021}).

\bibitem{openCARP_sw}
\bibinfo{author}{{openCARP Consortium}} \emph{et~al.}
\newblock \bibinfo{journal}{\bibinfo{title}{{openCARP} v11.0}}.
\newblock {\emph{\JournalTitle{RADAR4KIT}}} \url{10.35097/703}
  (\bibinfo{year}{2022}).

\bibitem{Fu-2013-ID14300}
\bibinfo{author}{{F}u, Z.}, \bibinfo{author}{{K}irby, R.~M.} \&
  \bibinfo{author}{{W}hitaker, R.~T.}
\newblock \bibinfo{journal}{\bibinfo{title}{{A} fast iterative method for
  solving the eikonal equation on tetrahedral domains}}.
\newblock {\emph{\JournalTitle{{SIAM} {J} {S}ci {C}omput}}}
  \textbf{\bibinfo{volume}{35}}, \bibinfo{pages}{c473--c494},
  \url{10.1137/120881956} (\bibinfo{year}{2013}).

\bibitem{Loewe-2019-ID12386}
\bibinfo{author}{{L}oewe, A.} \emph{et~al.}
\newblock \bibinfo{journal}{\bibinfo{title}{{P}atient-specific identification
  of atrial flutter vulnerability–a computational approach to reveal latent
  reentry pathways}}.
\newblock {\emph{\JournalTitle{{F}rontiers in {P}hysiology}}}
  \textbf{\bibinfo{volume}{9}}, \url{10.3389/fphys.2018.01910}
  (\bibinfo{year}{2019}).

\bibitem{Pilia-2021-ID15608}
\bibinfo{author}{{P}ilia, N.} \emph{et~al.}
\newblock \bibinfo{journal}{\bibinfo{title}{{ECG}deli - {A}n open source {ECG}
  delineation toolbox for {MATLAB}}}.
\newblock {\emph{\JournalTitle{{S}oftware{X}}}} \textbf{\bibinfo{volume}{13}},
  \bibinfo{pages}{100639}, \url{10.1016/j.softx.2020.100639}
  (\bibinfo{year}{2021}).

\bibitem{Kantelhardt_2003}
\bibinfo{author}{Kantelhardt, J.~W.}, \bibinfo{author}{Havlin, S.} \&
  \bibinfo{author}{Ivanov, P.~C.}
\newblock \bibinfo{journal}{\bibinfo{title}{Modeling transient correlations in
  heartbeat dynamics during sleep}}.
\newblock {\emph{\JournalTitle{Europhysics Letters ({EPL})}}}
  \textbf{\bibinfo{volume}{62}}, \bibinfo{pages}{147--153},
  \url{10.1209/epl/i2003-00332-7} (\bibinfo{year}{2003}).

\bibitem{Petrenas-2017-ID12969}
\bibinfo{author}{{P}etrenas, A.} \emph{et~al.}
\newblock \bibinfo{journal}{\bibinfo{title}{{E}lectrocardiogram modeling during
  paroxysmal atrial fibrillation: application to the detection of brief
  episodes.}}
\newblock {\emph{\JournalTitle{{P}hysiol {M}eas}}}
  \textbf{\bibinfo{volume}{38}}, \bibinfo{pages}{2058--2080},
  \url{10.1088/1361-6579/aa9153} (\bibinfo{year}{2017}).

\bibitem{PTB-XL-Plus}
\bibinfo{author}{Strodthoff, N.} \emph{et~al.}
\newblock \bibinfo{journal}{\bibinfo{title}{{PTB}-{XL}-{F}eat, a comprehensive
  electrocardiographic feature dataset}}.
\newblock {\emph{\JournalTitle{{in preparation}}}} .

\bibitem{nielsen15}
\bibinfo{author}{{N}ielsen, J.~B.} \emph{et~al.}
\newblock \bibinfo{journal}{\bibinfo{title}{{P}-wave duration and the risk of
  atrial fibrillation: Results from the {C}openhagen {ECG} study}}.
\newblock {\emph{\JournalTitle{Heart Rhythm}}} \textbf{\bibinfo{volume}{12}},
  \bibinfo{pages}{1887--1895}, \url{10.1016/j.hrthm.2015.04.026}
  (\bibinfo{year}{2015}).

\bibitem{unknown-0000-ID14904}
\bibinfo{author}{{N}agel, C.}, \bibinfo{author}{{P}ilia, N.},
  \bibinfo{author}{{L}oewe, A.} \& \bibinfo{author}{{D}össel, O.}
\newblock \bibinfo{journal}{\bibinfo{title}{{Q}uantification of interpatient
  12-lead {ECG} variabilities within a healthy cohort}}.
\newblock {\emph{\JournalTitle{{C}urrent {D}irections in {B}iomedical
  {E}ngineering}}} \textbf{\bibinfo{volume}{6}}, \bibinfo{pages}{493--496},
  \url{10.1515/cdbme-2020-3127} (\bibinfo{year}{2020}).

\bibitem{Bender-2022-ID18178}
\bibinfo{author}{{B}ender, J.} \emph{et~al.}
\newblock \bibinfo{journal}{\bibinfo{title}{{A} {L}arge-scale {V}irtual
  {P}atient {C}ohort to {S}tudy {ECG} {F}eatures of {I}nteratrial {C}onduction
  {B}lock}}.
\newblock {\emph{\JournalTitle{{C}urrent {D}irections in {B}iomedical
  {E}ngineering}}} \textbf{\bibinfo{volume}{8}}, \bibinfo{pages}{97--100},
  \url{10.1515/cdbme-2022-1026} (\bibinfo{year}{2022}).

\bibitem{strodthoff2020deep}
\bibinfo{author}{Strodthoff, N.}, \bibinfo{author}{Wagner, P.},
  \bibinfo{author}{Schaeffter, T.} \& \bibinfo{author}{Samek, W.}
\newblock \bibinfo{journal}{\bibinfo{title}{Deep learning for {ECG} analysis:
  Benchmarks and insights from {PTB-XL}}}.
\newblock {\emph{\JournalTitle{IEEE Journal of Biomedical and Health
  Informatics}}} \textbf{\bibinfo{volume}{25}}, \bibinfo{pages}{1519--1528}
  (\bibinfo{year}{2020}).

\bibitem{mehari2022self}
\bibinfo{author}{Mehari, T.} \& \bibinfo{author}{Strodthoff, N.}
\newblock \bibinfo{journal}{\bibinfo{title}{Self-supervised representation
  learning from 12-lead {ECG} data}}.
\newblock {\emph{\JournalTitle{Computers in Biology and Medicine}}}
  \textbf{\bibinfo{volume}{141}}, \bibinfo{pages}{105114}
  (\bibinfo{year}{2022}).

\bibitem{Dossel-2021-ID16522}
\bibinfo{author}{{D}össel, O.}, \bibinfo{author}{{L}uongo, G.},
  \bibinfo{author}{{N}agel, C.} \& \bibinfo{author}{{L}oewe, A.}
\newblock \bibinfo{journal}{\bibinfo{title}{{C}omputer modeling of the heart
  for {ECG} interpretation—a review}}.
\newblock {\emph{\JournalTitle{{H}earts}}} \textbf{\bibinfo{volume}{2}},
  \bibinfo{pages}{350--368}, \url{10.3390/hearts2030028}
  (\bibinfo{year}{2021}).

\bibitem{Luongo-2020-ID14752}
\bibinfo{author}{{L}uongo, G.} \emph{et~al.}
\newblock \bibinfo{title}{{A}utomatic {ECG}-based discrimination of 20 atrial
  flutter mechanisms: Influence of atrial and torso geometries}.
\newblock In \emph{\bibinfo{booktitle}{Computing in Cardiology}},
  vol.~\bibinfo{volume}{47}, \bibinfo{pages}{1--4},
  \url{10.22489/CinC.2020.066} (\bibinfo{publisher}{{IEEE}},
  \bibinfo{year}{2020}).

\bibitem{stenroos07}
\bibinfo{author}{{S}tenroos, M.}, \bibinfo{author}{{M}äntynen, V.} \&
  \bibinfo{author}{{N}enonen, J.}
\newblock \bibinfo{journal}{\bibinfo{title}{{A} {M}atlab library for solving
  quasi-static volume conduction problems using the boundary element method}}.
\newblock {\emph{\JournalTitle{Computer Methods and Programs in Biomedicine}}}
  \textbf{\bibinfo{volume}{88}}, \bibinfo{pages}{256--263}
  (\bibinfo{year}{2007}).

\bibitem{steffen_schuler_2022_7217554}
\bibinfo{author}{Schuler, S.} \& \bibinfo{author}{Loewe, A.}
\newblock \bibinfo{journal}{\bibinfo{title}{{FIM}\_{E}ikonal: v1.0}}.
\newblock {\emph{\JournalTitle{Zenodo}}} \url{10.5281/zenodo.7217554}
  (\bibinfo{year}{2022}).

\bibitem{ecg-syn}
\bibinfo{author}{Nagel, C.}, \bibinfo{author}{Eichhorn, N.} \&
  \bibinfo{author}{Loewe, A.}
\newblock \bibinfo{journal}{\bibinfo{title}{{ECG-Synthesization}: v1.0}}.
\newblock {\emph{\JournalTitle{Zenodo}}} \url{10.5281/zenodo.7293625}
  (\bibinfo{year}{2022}).

\bibitem{gillette2021automated}
\bibinfo{author}{Gillette, K.} \emph{et~al.}
\newblock \bibinfo{journal}{\bibinfo{title}{Automated framework for the
  inclusion of a his--purkinje system in cardiac digital twins of ventricular
  electrophysiology}}.
\newblock {\emph{\JournalTitle{Annals of Biomedical Engineering}}}
  \textbf{\bibinfo{volume}{49}}, \bibinfo{pages}{3143--3153}
  (\bibinfo{year}{2021}).

\bibitem{Odille-2017-ID12921}
\bibinfo{author}{{O}dille, F.}, \bibinfo{author}{{L}iu, S.},
  \bibinfo{author}{van {D}am, P.} \& \bibinfo{author}{{F}elblinger, J.}
\newblock \bibinfo{title}{{S}tatistical variations of heart orientation in
  healthy adults}.
\newblock In \emph{\bibinfo{booktitle}{{C}omputing in {C}ardiology {C}onference
  ({C}in{C})}}, vol.~\bibinfo{volume}{44}, \url{10.22489/CinC.2017.225-058}
  (\bibinfo{year}{2017}).

\bibitem{Loewe-2015-ID11872}
\bibinfo{author}{{L}oewe, A.} \emph{et~al.}
\newblock \bibinfo{title}{{L}eft and right atrial contribution to the {P}-wave
  in realistic computational models}.
\newblock In \bibinfo{editor}{van {A}ssen, H.}, \bibinfo{editor}{{B}ovendeerd,
  P.} \& \bibinfo{editor}{{D}elhaas, T.} (eds.)
  \emph{\bibinfo{booktitle}{{L}ecture {N}otes in {C}omputer {S}cience}}, vol.
  \bibinfo{volume}{9126} of \emph{\bibinfo{series}{{F}unctional {I}maging and
  {M}odeling of the {H}eart}}, \bibinfo{pages}{439–447},
  \url{10.1007/978-3-319-20309-6} (\bibinfo{year}{2015}).

\end{thebibliography}


\section*{Acknowledgements}

    This work was supported by the EMPIR programme co-financed by the participating states and from the European Union’s Horizon 2020 research and innovation programme under grant MedalCare 18HLT07. The authors also acknowledge the support of the British Heart Foundation Centre for Research Excellence Award III (RE/18/5/34216). SEW is supported by the British Heart Foundation (FS/20/26/34952).
    
       We thank the cardiologists Dr. Anna-Sophie Eberl, Dr. Ewald Kolesnik, 
    Dr. Martin Manninger-W\"unscher, Dr. Stefan Kurath-Koller, Dr. Susanne Prassl, 
    and Dr. Ursula Rohrer for their involvement in the clinical Turing tests and 
    for their feedback regarding the online platform and the ECG signal morphology. 
    We also thank Thomas Ebner and his colleagues from the Know-Center for the great 
    collaboration and the rapid implementation of our requirements in their 
    online platform TimeFuse.

\section*{Author contributions statement}

    All authors were involved in the writing and revision of the manuscript. \\
    K.G. built the ventricular-torso model cohort, parameterized and performed the simulations of the QRST complexes under both sinus and disease, conducted analysis on the clinical Turing tests, and organized the revision of the manuscript. \\
    M.G. managed the development of the testing platform for the clinical Turing test and developed tools to evaluate the test results. He also assisted in all aspects of simulations and model building.\\
    C.N. built the atrial model cohort, parameterized and performed the P~wave simulations under both sinus and disease conditions, designed and implemented the synthesization model, led the technical validation of simulated and clinical ECG biomarkers. \\
    J.B. performed and validated P~wave simulations for interatrial conduction block.\\
    B.W. ran simulations on the ventricular-torso model cohort, and extracted features from the clinical ECGs for technical validation. \\
    S.W. provided clinical insight and feedback on the clinical Turing tests. He also provided assistance on parameterization for both models. \\
    M.B. was involved in funding acquisition, provided guidance on relevant data processing and metrology aspects, reviewed and edited the final manuscript.\\
    T.S. provided motivation behind the study and gave clinical insight and guidance on relevant disease pathologies. \\
    O.D. was involved in funding acquisition and provided supervision of the atrial model cohort simulations.\\
    G.P was involved in funding acquisition and provided supervision of the ventricular-torso model cohort simulations. \\
    A.L. was involved in funding acquisition and provided supervision of the atrial model cohort simulations.\\
    
\section*{Competing Interests}

The authors declare no competing interests.

\clearpage
\section*{Figures \& Tables}





\begin{table}[ht]
\centering
\begin{tabular}{p{4cm}|p{3cm}|p{3cm}|p{1cm}|p{4cm}}
\multicolumn{5}{c}{Electrophysiological Parameters of QRS~Complex Simulations} \\
\hline
Entity & Parameter                               & Value            & Unit     & Reference \\
\hline
Geometry      & $\mathbf{\lambda}_{V,i}, i \in [1,13]$  & {[}1, 13{]}      & -        &  
Gillette~\etal~2021\cite{gillette2021framework}         \\ 
\cdashline{1-5}
Fascicular Sites       & $\vec{x}_{rv,mod}$   & \multirow{5}{2.2cm}{\{$\rho=0.2$,
$\phi=${[}0, 1.0{]},
$z=${[}0.1, 0.6{]},
$r=${[}0.2, 0.8{]},
$t=${[}0, 10{]}\}} & - &  \multirow{25}{5cm}{Durrer~\etal~1970\cite{durrer1970total}, \hspace{1cm} Gillette~\etal~2021\cite{gillette2021framework}}\\
& & & -  & \\
& & & -  & \\
& & & - & \\
& & & $\SI{}{\milli \second}$ & \\
 & $\vec{x}_{rv,sept}$   & \multirow{5}{2.2cm}{\{$\rho=0.8$,
$\phi=${[}-1.5, 1.5{]},
 $r=0.4$,
$z=${[}0.2, 0.4{]},
$t=10$ \}} & - & \\
& & & - & \\
& & & - & \\
 & & & - & \\
 & & & $\SI{}{\milli \second}$ & \\
 & $\vec{x}_{lv,sept}$   & \multirow{5}{2.2cm}{\{$\rho=0.2$,
$\phi=${[}-1.5, 1.5{]},
 $r=${[}0.05, 0.4{]},
$z=${[}0.3, 0.7{]},
$t=10$ \}} & - & \\
& & & - & \\
& & & - & \\
 & & & - & \\
  & & & $\SI{}{\milli \second}$ & \\
& $\vec{x}_{lv,ant}$   & \multirow{5}{2.2cm}{\{$\rho=0.2$,
$\phi=${[}1.0, $\pi${]},
$r=${[}0.05, 0.4{]},
$z=${[}0.2, 0.8{]},
$t=10$ \}} & - & \\
& & & - & \\
& & & - & \\
 & & & - & \\
  & & & $\SI{}{\milli \second}$ & \\
& $\vec{x}_{lv,post}$   & \multirow{5}{2.2cm}{\{$\rho=0.2$,
$\phi=${[}-$\pi$, -1.0{]},
$r=${[}0.05, 0.4{]},
$z=${[}0.2, 0.7{]},
$t=10$\}} & - & \\
& & & - & \\
& & & - & \\
 & & & - & \\
  & & & $\SI{}{\milli \second}$ & \\
\cdashline{1-5}
Conduction Velocity  & $cv_{endo}$  & 2.0    & $\SI{}{\meter \per \second}$ & Kassebaum~\etal~1966\cite{kassebaum1966electrophysiological}        \\
                 & $cv_{endo,r}$  & 1.0   & - & Gillette~\etal~2021\cite{gillette2021automated}         \\
                   & $cv_{myo}$  & 0.6    & $\SI{}{\meter \per \second}$ &    Taggart~\etal~2000\cite{taggart2000inhomogeneous}      \\
                 & $cv_{myo,r}$  & 4:2:1   & - & Taggart~\etal~2000\cite{taggart2000inhomogeneous}
         \\
\cdashline{1-5}
Myocardial Fiber    & $\alpha_{endo}$  & 60.0     & $^\circ$  & \multirow{4}{3cm}{Bayer~\etal~2012\cite{bayer2012novel}, Streeter~\etal~1969\cite{streeter1969fiber}}\\
Orientations        & $\alpha_{epi}$  & -60.0     &   $^\circ$        &   \\
                    & $\beta_{endo}$  & -65.0     &  $^\circ$         &   \\
                    & $\beta_{epi}$  & 25.0     &   $^\circ$        &   \\
\cdashline{1-5}
Heart Conductivity  & $\sigma_{il}$  & 0.34     & $\SI{}{\siemens \per \meter}$     &  \multirow{6}{3cm}{Roberts~\etal~1982\cite{roberts1982effect}}\\
                    & $\sigma_{in}$  & 0.06     &    $\SI{}{\siemens \per \meter}$       &   \\
                    & $\sigma_{it}$  & 0.06     &   $\SI{}{\siemens \per \meter}$         &   \\
                    & $\sigma_{el}$  & 0.12     &   $\SI{}{\siemens \per \meter}$         &   \\
                    & $\sigma_{en}$  & 0.08     &   $\SI{}{\siemens \per \meter}$         &   \\
                    & $\sigma_{et}$  & 0.08     &    $\SI{}{\siemens \per \meter}$        &   \\
\cdashline{1-5}
Volume-Conductor    & $\sigma_{torso}$  & 0.22   &   $\SI{}{\siemens \per \meter}$    &  \multirow{4}{3cm}{Keller~\etal~2010\cite{keller2010ranking}}\\ 
Conductivities      & $\sigma_{atria}$  & 0.0537 &   $\SI{}{\siemens \per \meter}$       &   \\
                    & $\sigma_{lungs}$  & 0.0389 &    $\SI{}{\siemens \per \meter}$       &   \\
                    & $\sigma_{blood}$  & 0.7    &    $\SI{}{\siemens \per \meter}$       &   \\
\hline
\end{tabular}
\caption{\label{tab:qrs_params} Model parameters for the electrophysiology within the ventricular simulations generating QRS simulations. Positioning, sizing, and timing of the 5 sites of fascicular breakthrough representing the His-Purkinje System within the ventricles provide variation in the QRS complex.  Fixed parameters were held constant at physiological values across all simulations as indicated.}          
\end{table}

\begin{table}[ht!]
\centering
\begin{tabular}{p{4cm}|p{3cm}|p{3cm}|p{1cm}|p{4cm}}
\multicolumn{5}{c}{Electrophysiological Parameters of T~Wave Simulations} \\
\hline
Entity & Parameter                               & Value            & Unit     & Reference \\
\hline
Ionic Model & $\vec{i}_{sinus}$  & \multirow{6}{2.2cm}{\{$V_{gate}=0.13$,
        $V_{min}=-86.2$,
        $V_{max}=40.0$,
        $\tau_{in}=0.3$,
        $\tau_{out}=5.4$,
        $\tau_{open}=80.0$\}} &  -   &  \multirow{6}{3cm}{Mitchell \& Schaeffer~2003\cite{mitchell2003two}}\\
        & & &  $\SI{}{\milli \volt}$& \\
& & & $\SI{}{\milli \volt}$ & \\
& & & -& \\
& & & - & \\
& & & - & \\ 
\cdashline{1-5}  
Repolarization Gradients & $APD_{min}$ & [150, 175]&  $\SI{}{\milli \second}$ & \multirow{6}{3cm}{Opthof~\etal~2017\cite{opthof2017cardiac}.Opthof~\etal~2016\cite{opthof2016dispersion},Keller~\etal~2011\cite{keller2011influence}} \\
 & $APD_{max}$ & [225, 250]&  $\SI{}{\milli \second}$ & \\
 & $\vec{q}_{w}$  &\multirow{4}{2.2cm}{\{$\rho=${[}-0.6, 0.0{]},
$\nu=${[}0.1, 0.15{]},
$\phi=0$,\\
$z=${[}0.9, 1.0{]}\}} & -  &  \\
& & & -  & \\
& &&  - & \\
& & & - & \\
\hline
\end{tabular}
\caption{ Model parameters for the electrophysiology within the ventricular simulations generating T~waves simulations. Base parameters of the action potential were held constant, but variations in action potential duration are prescribed using weighted gradients. }       
\label{tab:twave_params}
\end{table}

\begin{table}[ht]
\centering
\begin{tabular}{p{4cm}|p{3cm}|p{3cm}|p{1cm}|p{4cm}}
\multicolumn{5}{c}{Electrophysiological Parameters of Myocardial Infarction} \\
\hline
Entity & Parameter                               & Value            & Unit     & Reference \\
\hline
Sizing of Infarct & $d_{co}$  & {[}0, 1.0{]}   &  $\SI{}{\siemens \per \meter}$     &  Keller~\etal~2011\cite{keller2010ranking}\\ 
\cdashline{1-5}
Infarct Center      & $\vec{x}_{LAD,mi}$   & \multirow{3}{2.2cm}{\{
                $\phi=${[}0.0,2.0{]},
                $z=${[}0.1, 1.0{]}\}} 
                & - & \multirow{10}{2.2cm}{AHA~\etal~2002\cite{american2002standardized}} \\
                &  & & - &  \\
                &  & & - &  \\
 & $\vec{x}_{RAD,mi}$  & \multirow{3}{2.2cm}{\{$\phi=${[}-2.0, 0.0{]},
                $z=${[}0.2,1.0{]} \}} 
                & - &  \\
                &  & & - & \\
                &  & & - &  \\
 & $\vec{x}_{LCX,mi}$   & \multirow{4}{2.2cm}{\{ $\phi=${[}2.0, 3.14{]} $\cup$ {[}-3.14,-2.0{]},
                $z=${[}0.2, 1.0{]} \}} 
                &-  &   \\
                &  & & - & \\
                &  & & - &  \\
& & & - & \\
\cdashline{1-5}
Infarct Transmurality      & $\rho_{\nu,mi}$   & $\{0.3, 1.0\}$ & - & \\
\cdashline{1-5}
Conduction Velocity & $cv_{BZ}$  & 0.15 &     $\SI{}{\meter \per \second}$     &  \multirow{2}{3cm}{Mendonca~\etal~2018\cite{mendonca2018modeling}} \\
         & $cv_{BZ,r}$  & 1.0 &   -       &  \\
                   \cdashline{1-5}
Mitchell Schaeffer & $\vec{i}_{BZ}$  & \multirow{6}{2.2cm}{\{$V_{gate}=0.13$,
                $V_{min}=-73.1$,
                $V_{max}=12.5$,
                $\tau_{in}=0.45$,
                $\tau_{out}=3.6$,
                $\tau_{open}=44.0$\}} &  -        & \multirow{6}{3cm}{Mitchell,Schaeffer~2003\cite{mitchell2003two}, Loewe~\etal~2018\cite{loewe18}}  \\
Ionic Model & & & $\SI{}{\milli \volt}$ & \\
& & & $\SI{}{\milli \volt}$ & \\
& & & - & \\
& & & - & \\
& & & - & \\
\hline
\end{tabular}
\caption{\label{tab:mi_params}
Additional parameters were included to define infarct zones within the ventricular-torso model. Variations in the locations of the occlusion of the 3 primary arteries (LCA, LCX, and RCA) are based on clinical observations. Two different transmuralities are modeled. Fixed parameters comprise conductivity, conduction velocity, and the cellular settings. }
\end{table}

\begin{table}[ht]
\centering
\begin{tabular}{p{4cm}|p{3cm}|p{3cm}|p{1cm}|p{4cm}}
\multicolumn{5}{c}{Electrophysiological Parameters of P~wave simulations} \\
\hline
Entity                & Parameter                               & Value            & Unit     & Reference \\ \hline
Geometry       & $\mathbf{\lambda}_{A,i}, i \in [1,24]$  & {[}-3, 3{]}      & -        &  Nagel~\etal~2021\cite{Nagel-2021-ID16581}         \\ 
        & $\mathbf{\lambda}_{T,i}, i \in [1,2]$   & {[}-2, 2{]}      & -        &   Pishchulin~\etal~2017\cite{Pishchulin-2017-ID12226}        \\ 
\cdashline{1-5}
Atrial rotation       & $\alpha_x$                              & {[}-20, 20{]}    & $^\circ$ &  \multirow{3}{3cm}{Odille~\etal~2017\cite{Odille-2017-ID12921} }        \\
                      & $\alpha_y$                              & {[}-20, 20{]}    & $^\circ$ &          \\
                      & $\alpha_z$                              & {[}-20, 20{]}    & $^\circ$ &       \\ 
\cdashline{1-5}        
Atrial translation    & $t_x$                                   & {[}-10, 10{]}    & mm       &   \multirow{3}{3cm}{Odille~\etal~2017\cite{Odille-2017-ID12921}}         \\
                      & $t_y$                                   & {[}-10, 10{]}    & mm       &          \\
                      & $t_z$                                   & {[}-10, 10{]}    & mm       &           \\ 
\cdashline{1-5}
Transversal Conduction  & $\mathrm{CV_{bulk~tissue}}$             & {[}0.57, 0.85{]} & $\SI{}{\meter \per \second}$      &   \multirow{5}{3cm}{Loewe~\etal~2015\cite{Loewe-2015-ID11872}}          \\
Velocities       & $\mathrm{CV_{interatrial~connections}}$ & {[}0.46, 0.70{]} & $\SI{}{\meter \per \second}$      &           \\
                      & $\mathrm{CV_{crista~terminalis}}$       & {[}0.57, 0.85{]} & $\SI{}{\meter \per \second}$      &            \\
                      & $\mathrm{CV_{pectinate~muscles}}$       & {[}0.62, 0.92{]} & $\SI{}{\meter \per \second}$      &            \\
                      & $\mathrm{CV_{inferior~isthmus}}$        & {[}0.57, 0.85{]} & $\SI{}{\meter \per \second}$      &            \\ 
\cdashline{1-5}
Anisotropy ratios     & $\mathrm{AR_{bulk~tissue}}$             & 1.94             & -        &  \multirow{5}{3cm}{Loewe~\etal~2015\cite{Loewe-2015-ID11872}}          \\
                      & $\mathrm{AR_{interatrial~connections}}$ & 3                & -        &            \\
                      & $\mathrm{AR_{crista~terminalis}}$       & 2.56             & -        &         \\
                      & $\mathrm{AR_{pectinate~muscles}}$       & 3.24             & -        &          \\
                      & $\mathrm{AR_{inferior~isthmus}}$        & 1                & -        &         \\
\cdashline{1-5}                    
Torso conductivity    & $\sigma_{torso}$                              & 0.22             & $\SI{}{\siemens \per \meter}$      & Keller~\etal~2010\cite{keller2010ranking}          \\ 
\hline
\end{tabular}
\caption{\label{tab:atria_params} Model parameters for atrial simulations. Values were varied randomly following a uniform distribution in the specified intervals. Fixed parameters comprise anisotropy ratios and torso conductivity, which were defined as indicated in the respective column.}
\end{table}

\FloatBarrier

\end{document}